\documentclass[11pt]{article}
\usepackage{epsfig}
\usepackage{verbatim}
\usepackage{graphicx}
\usepackage{amsfonts}
\usepackage{amsmath}
\usepackage{amssymb}
\usepackage{amsthm}
\usepackage{newlfont}
\usepackage{natbib}
\usepackage{setspace}
\usepackage{longtable}
\usepackage{color,xcolor}
\usepackage{algorithmic}
\usepackage{algorithm}
\usepackage{flowchart}
\usepackage{tikz}
\usetikzlibrary{arrows,shapes,chains}
\makeatletter 
\def\singlespace{\def\baselinestretch{1}\@normalsize}

\theoremstyle{definition}
\newtheorem{thm}{Theorem}[]

\newtheorem{rem}{Remark}

\begin{document}
\title{Individual Heterogeneity Learning in Distributional Data Response Additive Models}
\author{Zixuan Han, Tao Li, Jinhong You \\[2ex]
School of Statistics and Management, Shanghai University of Finance and Economics
}
	
\date{}
	
\maketitle

%%%%%%%%%%%%%%%%%%%%%%%%%%%%%%%%%%%%%%%%%%%%%%%%%%%%%%%%%%%%%%%%%%%%%%%%%%%%%%%%%%%%
%%%%%%%%%%%%%%%%%%%%%%%%%%%%%%%%%%%%%%%%%%%%%%%%%%%%%%%%%%%%%%%%%%%%%%%%%%%%%%%%%%%%

\begin{abstract}
In many complex applications, data heterogeneity and homogeneity exist simultaneously. Ignoring either one will result in incorrect statistical inference. In addition, coping with complex data that are non-Euclidean becomes more common. To address these issues we consider a distributional data response additive model in which the response is a distributional density function and the individual effect curves are homogeneous within a group but heterogeneous across groups, the covariates capturing the variation share common additive bivariate functions. A transformation approach is first utilized to map density functions into a linear space. We then apply the B-spline series approximating method to estimate the unknown subject-specific and additive bivariate functions, and identify the latent group structures by hierarchical agglomerative clustering (HAC) algorithm. Our method is demonstrated to identify the true latent group structures with probability approaching one. To improve the efficiency, we further construct the backfitted local linear estimators for grouped structures and additive bivariate functions in post-grouping model. We establish the asymptotic properties of the resultant estimators including the convergence rates, asymptotic distributions and the post-grouping oracle efficiency. The performance of the proposed method is illustrated by simulation studies and empirical analysis with some interesting results.
\end{abstract}

\noindent%

%%%%%%%%%%%%%%%%%%%%%%%%%%%%%%%%%%%%%%%%%%%%%%%%%%%%%%%%%%%%%%%%%%%%%%%%%%%%%%%%%%%%%%%%%%%%%%%%%%%%%%%%%%%%%%%%%%

%%%%%%%%%%%%%%%%%%%%%%%%%%%%%%%%%%%%%%%%%%%%%%%%%%%%%%%%%%%%%%%%%%%%%%%%%%%%%%%%%%%%%%%%%%%%%%%%%%%%%%%%%%%%%%%%%%
\section{Introduction}\label{sec:intro}

With the development of modern technology, data are increasingly being measured and recorded at several discrete times or a continuous time interval, which are called functional data, and have become a prevailing type of data \citep{r82}. Functional data analysis provides statistical methodology to deal this kind of data, among which functional regression is widely used to model the relationship of responses and predictors. A great quantity of researches focus on this field, which can be divided into three classes according to whether the responses or predictors are functional or scalar data as follows. The first case is that both responses and predictors are functional data, see \citep{mr03, ymw05, hmwy10, jw11, chl16}; the second one is functional responses with scalar predictors, see \citep{fy03, fy05, hh07, hmv13}; while the other one is scalar responses with functional predictors, see \citep{zlk12, lzz16, lhz17}.

As a specific case of functional data, data consist of samples of distributions or densities appear in various research domains increasingly often. Examples giving rise to such data are distributions of cross-sectional or intraday stock return \citep{sm15, kmps2019}, mortality densities \citep{pm19}, distribution of intra-hub connectivity in neuroimaging \citep{setal16, pcm19}. Compared with conventional functional data, distribution or density function takes data as a whole to present its internal structure without the limitation of sample order and the information dimension. In recent years, along with the application of such data, the attention has turn to the complex regression models in which the random distributions or probability densities serve as responses or predictors. In this article, we focus on the model with density functions as responses coupled with scalar predictors, viz., distribution-on-scalar regression model. 

Considering the density functions as elements of a Hilbert space, they do not constitute a linear subspace due to the inherent constraints of being nonnegative and integrated to one. To deal with this constrain, one way is to adopt the geometric approach by choosing a suitable metric. With the the infinite-dimension version of Aitchison geometry, \cite{tmmhf18} defined a density-on-scalar linear regression model in Bayes Hilbert spaces. \cite{clm20} proposed distribution-on-distribution regression by adopting Wasserstein metric and tangent bundle of the Wasserstein space. Except this, some other models are within the broad framework of non-Euclidean data. For instance, \cite{pm19} proposed the Fr\'echet linear regression in a general metric space equipped with the Wasserstein metric. \cite{jk2020a} developed a unified nonparametric additive regression models with Hilbertian responses where density-valued variables constitute Bayes-Hilbert spaces equipped with a suitable inner product. Another way of dealing with the nonlinear constrain of density functions is to map them into Hilbert space by transformation method. \cite{pm16} proposed a continuous and invertible transformation, e.g., the log quantile density transformation (LQD) to map probability densities to an unrestricted space of square integrable functions for further modeling and statistical inferences. 

In conventional statistical analysis, data is generally assumed to be  homogeneous. However, this assumption might be inappropriate in many practical applications when the data are collected from objects with different characteristics or in different situations. Ample empirical studies show that inter-class individual homogeneity and intra-class heterogeneity may exist simultaneously, while ignoring the individual heterogeneity during the analysis may lead to incorrect statistical results, and ignoring the homogeneity will result in inefficient statistical inference. Therefore, the density functions within a heterogeneous population should be clustered into several homogeneous groups by some classification measures.

A mature literature proposes various methods to identify the latent group structures. \cite{vl17} applied a distance-based clustering algorithm to the kernel estimation of nonparametric regression function. \cite{vl18} extended it to a multiscale statistic to avoid the selection of specific bandwidth. \cite{ssp16} developed a so-called classifier-Lasso (C-Lasso) shrinkage method to the linear panel structure model. As a extension, \cite{swj19} developed a penalized-sieve-estimation-based C-Lasso procedure to heterogeneous time-varying panel data. \cite{c19} proposed a kernel based hierarchical agglomerative clustering (HAC) algorithm  with less restrictive assumptions to the same kind data compared with former method. Relevant approaches discussed above are contributed in the context of  functional or panel data.

The focus of this article is to identify and estimate the latent group structures in additive regression model with density responses. The heterogeneity in the density responses are indeed found when we explore the COVID-19 data. On February 11, 2020, the disease caused by severe acute respiratory syndrome coronavirus 2 (SARS-CoV-2) was officially named as `COVID-19' by the World Health Organization (WHO). With the increase in the infection range and transmission speed, the WHO declared the COVID-19 a pandemic on March 11, 2020. At that time, there were about 37371 confirmed cases and 1130 deaths reported in about 114 countries. By December 15, 2020, according to the official repository of WHO (https://covid19.who.int), there have been 73,315,291 confirmed cases reported in about 271 countries and regions, leading to about 1,627,480 deaths. The tremendous increase on these numbers indicate that the COVID-19 is impacting the whole world drastically. It should be noted that due to the medical conditions and adopted prevention policies, along with some other factors both subjective and objective, the epidemic situation in different countries may vary to some extent.
 
\begin{figure}[!ht]\centering
	\includegraphics[width=8cm]{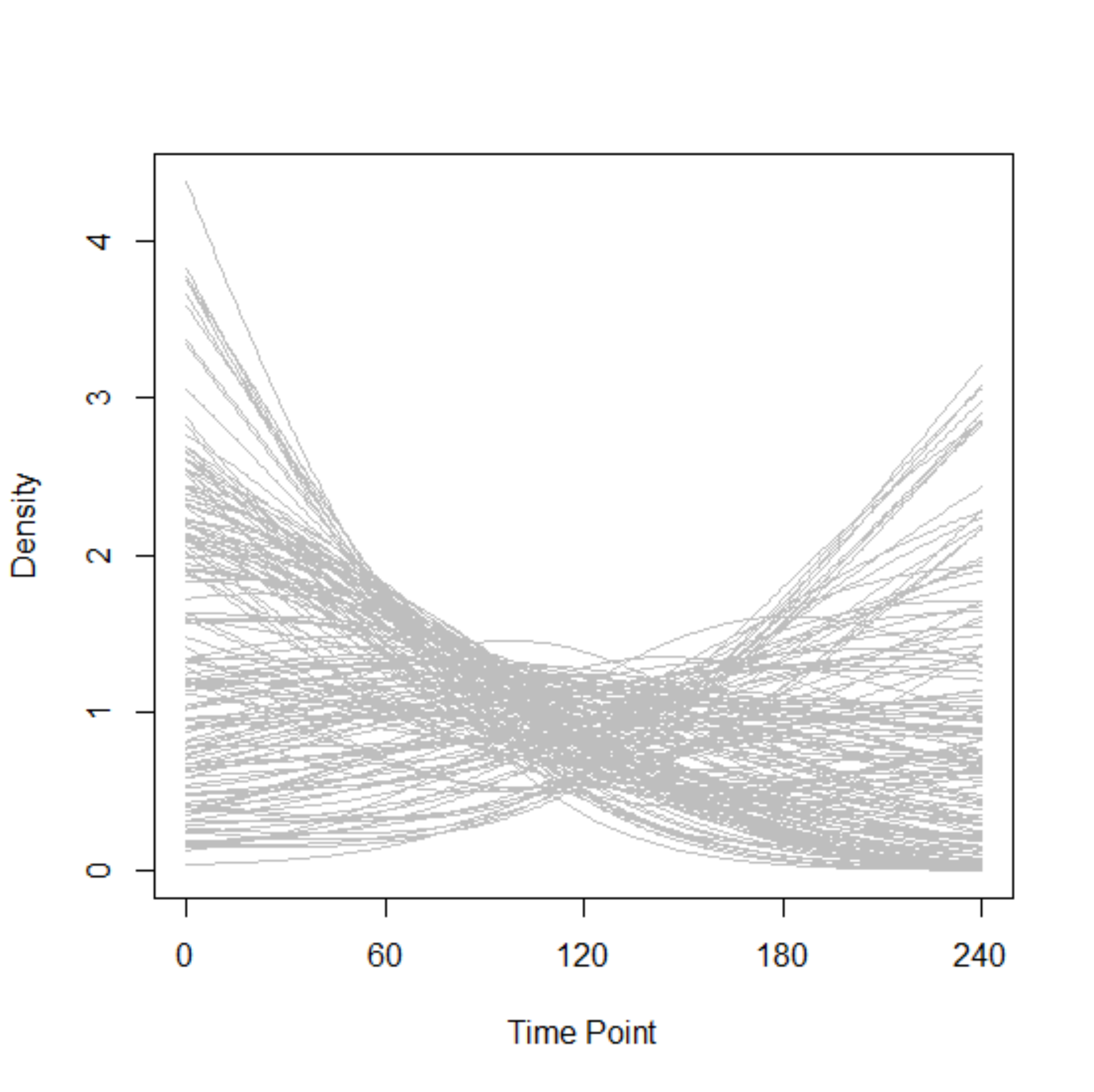}
	\caption{Densities of relative daily mortality rate of COVID-19 in 149 countries over a period of 240 days. } \label{fg:covid}
\end{figure}

To explore the epidemic in each country relative to the global situation with influential factors, we take the density of relative daily mortality rate over a period 240 days in each country as response, where the daily mortality rate is defined as the ratio of deaths per day to the total population of each country, and the relative one is defined as the ratio of daily mortality rate in each country to the total mortality rate of all countries in the world. The predictors considered for explanation are `aging' (the percentage of population ages 65 and above), `beds' (the number of hospital beds per 1000 people), `physicians' (the number of physicians per 1000 people), `nurses' (the number of nurses per 1000 people) , `GDP' (the GDP per capita in the US dollar) , and `diabetes' (the percentage of population with diabetes). Since the outbreak time varies in different countries, we set the date on which each country first reported at least 30 deaths as the origin of time scale. Totally 149 countries are taken into account and the densities of daily mortality rate for these 149 countries are presented by Figure \ref{fg:covid}. The existence of different shapes of the densities among countries intrigues us to consider the subject-specific functions and impose latent grouped structures representing heterogeneity in the functional additive model proposed by \cite{hmp19}, i.e.
\begin{equation*}
\Psi(z_{i})(u)=g_{i,0}(u)+\sum_{l=1}^{6}g_{l}(u,X_{i,l})+\varepsilon_i(u),  \quad 1\leq i \leq 149,\ 1\leq l \leq 6,
\end{equation*}
where $z_i$ is the density of the relative daily mortality rate over a period 240 days in country $i$, $\Psi(\cdot)$ is the LQD transformation, $ \boldsymbol{X_{i}}=(X_{i,1},...,X_{i,6})^{\tau} $ is the covariate vector of country $ i $. Specifically, $ g_{i,0}(u)=m_{k|K,0}(u) $ if country $ i $ is sharing the kind $ k $ pattern, and $ m_{\cdot|K,0}(u) $ is one of $ K $ group-specific functions. 

\begin{figure}[!ht]\centering
	\includegraphics[width=16.5cm]{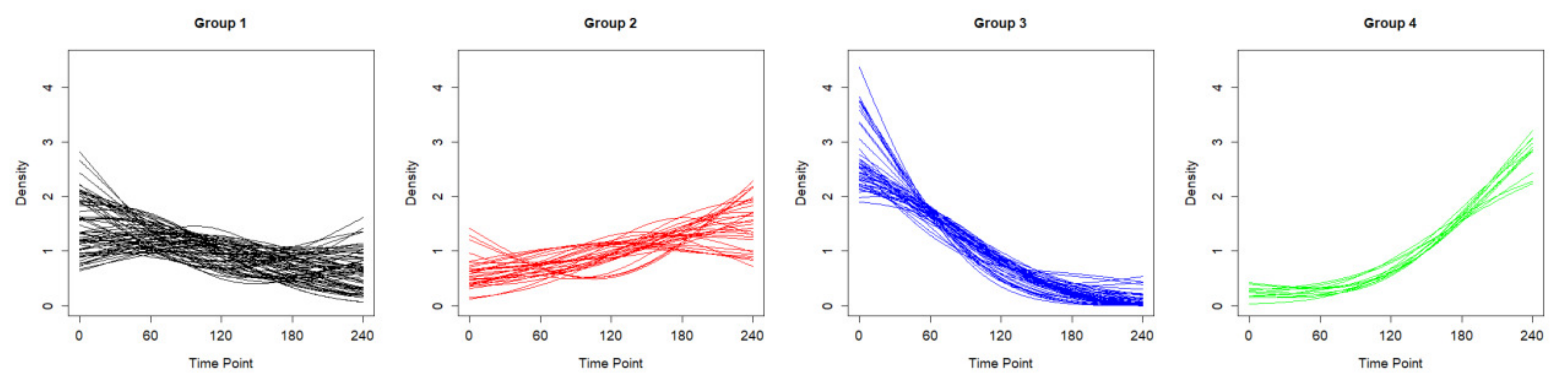}
	\caption{Densities of relative daily mortality rate of COVID-19 over a period of 240 days in four different groups corresponding to the clustering results. } \label{fg:covid-gf}
\end{figure}

Motivated by \cite{c19}, we apply HAC method to the estimation of individual functions $ g_{i,0}(u) $, based on which the density functions of relative daily mortality rate can be clustered into four groups with different patterns of epidemic situations, i.e., $ K=4 $, which is shown in Figure \ref{fg:covid-gf}. The figure is also served as sufficient and necessary evidence that the functional additive model implemented for the analysis of COVID-19 data should consider the individual heterogeneity among countries.

In order to incorporate both inter-class homogeneity and intra-class heterogeneity in the data, we extend the additive functional regression for densities as responses proposed by \cite{hmp19} to the case of density functions with heterogeneity: 
\begin{equation}\label{eq:model}
f_{i}(u)=\Psi(z_{i})(u)=g_{i,0}(u)+\sum_{l=1}^{p}g_{l}(u,x_{i,l})+\varepsilon_i(u), \quad 1\leq i\leq n,
\end{equation}
and
\begin{equation}\label{eq:latentstructure}
g_{i,0}(u)=
\begin{cases}
m_{1|K,0}(u), \quad  & i\in G_{1},\\
m_{2|K,0}(u), \quad  & i\in G_{2},\\
\cdots &\cdots \\
m_{K|K,0}(u), \quad  & i\in G_{K}.
\end{cases}
\end{equation}
where $\{ G_{1},...,G_{K}\} $ is a partition of the index set $\{1,2,\cdots,n\}$, which means that $ \bigcup\limits_{k=1}^{K} G_{k}=\{1,2,...,n\} $, and $ G_{i}\cap G_{j}=\emptyset $, $ ||m_{i|K,0}(\cdot)-m_{j|K,0}(\cdot)||_{2} \neq 0 $ for any $ i\neq j $. We assume that the number of groups $ K $ and the membership in each individual group are unknown.

In the above model, $ z_{i}(u)\in \mathcal{F} $ are the random densities, each coupled with the $ p- $dimensional covariates $ \boldsymbol{x}_{i}=(x_{i,1},\cdots,x_{i,p})^{\tau} $, with common support $ S_{x} $. Without loss of generality, we take $ S_{x}= [0,1] $. Denote $ \Psi(\cdot) $ as the log quantile density transformation, and $ f_{i}=\Psi(z_{i}) $.  $ g_{i,0}(\cdot) $ is the subject-specific function and $ g_{l}(\cdot,x_{l}) $ is the bivariate additive components. For the purpose of the identification, we assume that $ g_{l}(\cdot,x_{l}) $ are: $ E[g_{l}(u,x_{i,l})]=0 $, for $u\in[0,1]$, $i=1,...,n $. $ \{\boldsymbol{\varepsilon}_i \}_{i=1}^n$ are independent processes with $ E(\varepsilon_i(u)|\boldsymbol{x}_{i})=0$ and $ Cov(\varepsilon_i(u)|\boldsymbol{x}_{i})=\sigma^{2}_{i}(u)$.

Obviously, our proposed model \eqref{eq:model} reveals that it is a natural extension of functional additive model. Specifically, if the subject-specific functions are homogeneous, viz., $ K=1 $, $g_{i,0}(\cdot)\equiv g_{0}(\cdot)$, $i=1,\cdots,n$, then the model (\ref{eq:model}) becomes the additive function regression for densities as responses  proposed by \cite{hmp19}
\begin{equation*}\label{eq:homoregression}
f_{i}(u)=\Psi(z_{i})(u)=g_{0}(u)+\sum_{l=1}^{p}g_{l}(u,x_{i,l})+\varepsilon_i(u),  \quad 1\leq i \leq n,\ 1\leq l \leq p. 
\end{equation*}

Actually, \cite{hmp19} has conducted great work and gained excellent academic achievements. It proposed a novel additive functional regression model to accommodate random densities as responses and multivariate predictors with the increasing popularity of analyzing data in the form of distributional functions, and established relevant theoretical properties and statistical inference. Motivated by this model, we are interested to identify the subgroups when analyzing data consisting of densities with heterogeneity. 

In real applications, since only a random sample generated by the density is available, we first estimate each density by the modified kernel density approach proposed by \cite{pm16} along with the LQD transformation. The identification and estimation methodology of the latent group structures are accomplished in three steps. First, we utilize B-spline series approximating approach to estimate subject-specific function $g_{i,0}(u)$ and bivariate additive components $g_l(u,x_l)$. Then, we apply HAC method to identify the membership of each latent group. Finally, backfitted local linear regression is applied to improve the efficiency of group-specific function $m_{k|K,0}(u)$ and $g_l(u,x_l)$. Theoretical results concerning the resultant estimations including the uniform convergence rate of the initial estimation, the consistency of the estimation of group number and partitions, both the uniform convergence rate and the asymptotic normality of the group-specific functions and the post-clustering estimation of additive component are also established.

The rest of this article is organized as follows. In Section 2, the modified kernel estimation and the LQD transformation for density functions are introduced as the preliminary works. Section 3 presents the procedure for identification and estimation of latent group structures and additive components in the model. Theoretical results are included in Section 4. The Monte Carlo simulations are conducted to illustrate the efficiency of proposed procedure in Section 5. We also demonstrate the application of proposed method by analyzing the COVID-19 and GDP data in Section 6. A discussion follows in Section 7. Detailed proofs of theoretical results and additional numeric results are in the supplementary materials.

%%%%%%%%%%%%%%%%%%%%%%%%%%%%%%%%%%%%%%%%%%%%%%%%%%%%%%%%%%%%%%%%%%%%%%%%%%%%%%%%%%%%%%%%%%%%%%%%%%%%%%%%%%%%%%%%%%
\section{Preliminaries}

In this section, some preliminaries required for the model will be presented. 
%%%%%%%%%%%%%%%%%%%%%%%%%%%%%%%%%%%%%%%%%%%%%%%%%%%%%%%%%%%%%%%%%%%%%%%%%%%%%%%%%%%%%%%%%%%%%%%%%%%%%%%%%%%%%%%%%%
\subsection{Log Quantile Density Transformation}

Let $ \mathcal{F} $ be the class of univariate continuous probability density function $ z(u) $, with a common support $S$ and satisfying  $\int_{\mathcal{R}}^{}u^{2}z(u)du<\infty $. Without loss of generality, we take $ S= [0,1] $. For each $z(x)\in \mathcal{F}$,  let $ F(x)=\int_{-\infty}^{x}z(u)du $, $ Q(u) $ be the corresponding quantile function, i.e, $ Q=F^{-1} $, and $ q(u) $ be the quantile density function defined as the derivative of quantile function, i.e. $ q(u)=Q^{'}(u)=\frac{d}{du}F^{-1}(u) $ for $ u\in [0,1] $. 

To map each $z(u)\in \mathcal{F}$ into the linear space $L^2([0,1])$, we utilize the log quantile density transformation \citep{pm16} that is defined as
\begin{equation*}
\Psi(z)(u)=log(q(u))=-log\{z(Q(u))\}, \quad u\in[0,1].
\end{equation*}

%%%%%%%%%%%%%%%%%%%%%%%%%%%%%%%%%%%%%%%%%%%%%%%%%%%%%%%%%%%%%%%%%%%%%%%%%%%%%%%%%%%%%%%%%%%%%%%%%%%%%%%%%%%%%%%%%%
\subsection{Modified Density Estimation}

A challenge for fitting the regression model with density response is that in real applications, $z_i(u)$, thus $f_i(u)$, is unobservable, and can only be estimated from the random sample generated by $z_i(u)$, viz., $ Y_{i1},...,Y_{iT_{i}}\stackrel{iid}{\sim} z_i(u)$, with corresponding covariate $ \boldsymbol{X_{i}}=(X_{i,1},...,X_{i,p})^{\tau} $. Without loss of generality, we assume $ T_{i}=T $. 
Due to the boundary effects of conventional kernel density estimators, \cite{pm16} proposed a modified kernel density estimator $\hat{z}_i(u)$ as follows, i.e.
\begin{equation*}
\hat z_i(u)=\sum_{l=1}^{T}\mathcal{K}\left(\dfrac{u-Y_{il}}{h}\right)w(u,h)\biggl/\sum_{l=1}^{T}\int_{0}^{1}\mathcal{K}\left(\dfrac{y-Y_{il}}{h}\right)w(y,h)dy,\qquad i=1,2,\cdots,n,
\end{equation*}
with the weight function $ w(u,h) $ be $ \left(\int_{-u/h}^{1}\mathcal{K}(v)dv\right)^{-1} $ as $ u\in[0,h) $, $ \left(\int_{-u/h}^{1}\mathcal{K}(v)dv\right)^{-1} $ as $ u\in(1-h,1] $ and 1 otherwise. Bandwidth $h < 1/2$, $ \mathcal{K} $ is of bounded variation and symmetric of 0, satisfying the conditions that $ \int_{0}^{1}\mathcal{K}(u)du>0 $, $ \int_{\mathcal{R}}^{}|u|\mathcal{K}(u)du $, $ \int_{\mathcal{R}}^{}\mathcal{K}^{2}(u)du $ and $ \int_{\mathcal{R}}^{}|u|\mathcal{K}^{2}(u)du $ are finite. Different from the conventional kernel density estimator, modified kernel density estimator $\hat{z}_i(u)$ possesses the consistency property, viz.,
\[\sup_{z_i(u)\in \mathcal{F} }||\hat{z}_i(u)-z_i(u)||_{\infty}\,\rightarrow\, 0, \hspace{1cm} \text{as }\,T\,\rightarrow\,\infty.\]

Based on the estimation of density functions, to fit model \eqref{eq:model}, we take $\hat{z}_i(u)$ as the substitution of $z_i(u)$ and denote $\tilde{f}_i(u)=\Psi(\hat{z}_i)(u)$, then the model \eqref{eq:model} is written as 
\begin{equation}\label{eq:model2}
\tilde{f}_i(u)=g_{i,0}(u)+\sum_{l=1}^{p}g_{l}(u,x_{i,l})+\varepsilon_{f_i}(u)+\varepsilon_i(u),  \quad 1\leq i \leq n,\ 1\leq l \leq p,
\end{equation}
where $\varepsilon_{f_i}(u)=\tilde{f}_i(u)-f_i(u)$ is the random error resulting from the estimation of $f_i(u)$.

%%%%%%%%%%%%%%%%%%%%%%%%%%%%%%%%%%%%%%%%%%%%%%%%%%%%%%%%%%%%%%%%%%%%%%%%%%%%%%%%%%%%%%%%%%%%%%%%%%%%%%%%%%%%%%%%%%

\section{Identification and estimation methodology}

In this section, we provide the methodology for identification of the latent group structures and estimation of additive components in the proposed model through a three-step procedure. The identification and estimation procedure is provided in Algorithm 1.

%%%%%%%%%%%%%%%%%%%%%%%%%%%%%%%%%%%%%%%%%%%%%%%%%%%%%%%%%%%%%%%%%%%%%%%%%%%%%%%%%%%%%%%%%%%%%%%%%%%%%%%%%%%%%%%%%%

\begin{algorithm}[H]
\caption{Identifying Heterogeneity in Additive Model with Density Responses}

\vspace{0.2cm}
\setstretch{1.2}
\begin{algorithmic}[1]
\STATE \textbf{Data:} $ (Y_{it}, \boldsymbol{X_{i}}) $,  where $ \boldsymbol{X_{i}}=(X_{1},...,X_{p})^{\tau}$, $ t=1,...,T_{i}, \ i=1,...,n, $

\STATE \textbf{Estimated Density:} Modified kernel density estimation $ \hat{z}_{i}$.
	
\STATE \textbf{Transformed Density:} $ \hat{f}_{i}=\Psi(\hat{z}_{i}). $
	
\STATE \textbf{Initial Estimation:} B-spline approximation
$ \hat{g}_{i,0}(u), \ \hat{g}_{l}(u,x_{l}),  \ l=1,...p, \ i=1,...,n. $ 
	
\STATE \textbf{HAC Algorithm:} Estimated group sets index $\{\hat{G}_{1},...,\hat{G}_{K}\}.$ 

\begin{itemize}
\setstretch{0.8}
\item Begin with $ n $ groups, where each subject is a group.
\item Merge the two groups with smallest distance into one.
\item Recalculate the distance between new groups after each grouping.
\item Repeat the previous two steps until the number of groups achieves $ K $.
\end{itemize}
	
\STATE \textbf{Refined Estimation:}  Backfitted local linear estimation $ \hat{m}_{k|K,0}(u), \ \tilde{g}_(u,x_{l})$, $ k=1,...,K$, $ l=1,...,p$. 

\end{algorithmic}
\end{algorithm}

%%%%%%%%%%%%%%%%%%%%%%%%%%%%%%%%%%%%%%%%%%%%%%%%%%%%%%%%%%%%%%%%%%%%%%%%%%%%%%%%%%%%%%%%%%%%%%%%%%%%%%%%%%%%%%%%%%

\subsection{Initial estimation}

Spline series approximating method is commonly used to estimate unknown nonparametric functions, with detailed practical guidance in \cite{db78} and \cite{s94}. 

Let $ \{B_1(u),B_1(u),...,B_{N_{0}+1}(u)\} $ be the set of B-spline basis functions of order $ q $ with $ L_{0} $ interior knots, where  $ N_{0}+1=L_{0}+q $. Let  $ \{B_{1,l}(x_{l}),..., B_{N_{l}+1,l}(x_{l})\} $ be the set of B-spline basis functions of order $ q $ for $ x_{l} $ ($ l=1,\cdots,p $) with $ L_{l} $ interior knots, where $ N_{l}+1=L_{l}+q $. Then, defined the normalized spline basis of $ x_{l} $ ($ l=1,\cdots,p $) as $ \{b^{*}_{1,l}(x_{l}),b^{*}_{2,l}(x_{l}),...,b^{*}_{N_{l},l}(x_{l})\} $, and denote $ b_{m}(u)=N_{0}^{1/2}B_{m}(u) $ as the scaled version of $ B_{m}(u) $. 

Based on the basis functions, define the tensor product of B-spline basis as
\begin{equation*}
 b_{m,k,l}(u,x_l)=b_{m}(u)b^{*}_{k,l}(x_l), \quad 1\leq m\leq N_{0}, \ 1\leq k\leq N_{l}, \ 1\leq l\leq p.
\end{equation*}

The spline approximation of subject-specific and bivariate components are given by
\begin{equation*}
g_{i,0}(u)\approx\sum_{j=1}^{N_{0}}\lambda_{i,j}b_{j}(u),\quad
g_{l}(u,x_{l})\approx\sum_{m=1}^{N_{0}}\sum_{k=1}^{N_{l}}\lambda_{m,k,l}b_{m,k,l}(u,x_l), \quad 1\leq l\leq p, \quad 1\leq i\leq n.
\end{equation*}

Denote $ g_{i}(u,\boldsymbol{x})=g_{i,0}(u)+\sum_{l=1}^{p}g_{l}(u,x_{l}) $,
therefore, the model \eqref{eq:model} can be written as $ f_{i}(u)=g_{i}(u,\boldsymbol{x})+\varepsilon_{i}(u) $, and we can approximate $ g_{i}(u,\boldsymbol{x}) $ by
\begin{equation*}
g_{i}(u,\boldsymbol{x})\approx\sum_{j=1}^{N_{0}}\lambda_{i,j}b_{j}(u)+\sum_{l=1}^{p}\sum_{m=1}^{N_{0}}\sum_{k=1}^{N_{l}}\lambda_{m,k,l}b_{m,k,l}(u,x_{l}), \quad 1\leq l\leq p, \ 1\leq i\leq n.
\end{equation*}

Then, the corresponding estimations are
\begin{equation}\label{eq:initial}
\hat{g}_{i,0}(u)=\sum_{j=1}^{N_{0}}\hat{\lambda}_{i,j}b_{j}(u),\quad
\hat{g}_{l}(u,x_{l})=\sum_{m=1}^{N_{0}}\sum_{k=1}^{N_{l}}\hat{\lambda}_{m,k,l}b_{m,k,l}(u,x_{l}),	
\end{equation}
\begin{equation*}
\hat{g}_{i}(u,\boldsymbol{x})=\sum_{j=1}^{N_{0}}\hat{\lambda}_{i,j}b_{j}(u)+\sum_{l=1}^{p}\sum_{m=1}^{N_{0}}
\sum_{k=1}^{N_{l}}\hat{\lambda}_{m,k,l}b_{m,k,l}(u,x_{l}),
\end{equation*}
where $ \boldsymbol{\hat{\lambda}}=(\hat{\lambda}_{1,1},...,\hat{\lambda}_{n,N_{0}},
\hat{\lambda}_{1,1,1},...,\hat{\lambda}_{N_{0},N_{l},p})^{T} $ is a $ N_{0}(n+pN_{l})$-dimensional vector satisfying
\begin{equation}\label{eq:lambda1}
\boldsymbol{\hat{\lambda}}=\arg\min_{\boldsymbol{\lambda}}\sum_{i=1}^{n}\sum_{t=1}^{T}\bigg[f_{i}(u_{t})-\sum_{j=1}^{N_{0}}\lambda_{i,j}b_{j}(u_{t})
-\sum_{l=1}^{p}\sum_{m=1}^{N_{0}}\sum_{k=1}^{N_{l}}\lambda_{m,k,l}b_{m,k,l}(u_{t},X_{i,l})\bigg]^{2}.
\end{equation}

%%%%%%%%%%%%%%%%%%%%%%%%%%%%%%%%%%%%%%%%%%%%%%%%%%%%%%%%%%%%%%%%%%%%%%%%%%%%%%%%%%%%%%%%%%%%%%%%%%%%%%%%%%%%%%%%%%

\subsection{Identifying latent groups structures via HAC algorithm}
Based on the initial estimations of subject-specific functions \eqref{eq:initial}, the classic hierarchical agglomerative clustering (HAC) algorithm is applied for identifying the latent groups  $\{ G_{1},...,G_{K}\} $ given the group number $ K $. To tackle this problem, we need a metric to measure the distance between pair of functions $ g_{i,0}(\cdot) $ and $ g_{j,0}(\cdot) $. For instance, \cite{clwz19} used $L_1$-distance to measure the similarity of the coefficient functions in nonlinear models. \cite{vl17} combined $L_2$-distance with $k$-means procedure to obtain the group structure. \cite{vl18} developed the multiscale techniques based on $L_{\infty}$ for clustering. 

Similarly as \cite{cfs99}, in this article we work with the general $ L_{q}$-distance
\begin{equation*}
d_{i,j}=\int_{0}^{1}||g_{i,0}(u)-g_{j,0}(u)||_{q}du,
\end{equation*}
which can be estimated as
\begin{equation*}
\hat{d}_{i,j}=\frac{1}{T}\sum_{t=1}^{T}||\hat{g}_{i,0}(u_{t})-\hat{g}_{j,0}(u_{t})||_{q}.
\end{equation*}

Assume that the number of groups $ K $ is given. The HAC algorithm for finding out the latent group structures among the individual subject-specific functions can be summarized as follows. First, begin with $ n $ groups with each subject as a group. Second, calculate the distance matrix $ \hat{M}_{n\times n}=(\hat{d}_{i,j}) $, find the smallest element except for the main diagonal elements and merge the corresponding groups into one. Next,
recalculate the distance between new groups and the distance matrix with reduced size after each grouping. The distance between two groups is defined as the furthest distance between any two functions with one in a group and the other one in another group. Finally, repeat the previous two steps until the number of groups achieves $ K $.

%%%%%%%%%%%%%%%%%%%%%%%%%%%%%%%%%%%%%%%%%%%%%%%%%%%%%%%%%%%%%%%%%%%%%%%%%%%%%%%%%%%%%%%%%%%%%%%%%%%%%%%%%%%%%%%%%%
\subsection{Estimation of Latent Group Structures}

Given the true group membership, the model (\ref{eq:model}) can be written in group-structure form
\begin{equation}\label{eq:latent-model}
f_{i}(u)=m_{k|K,0}(u)+\sum_{l=1}^{p}g_{l}(u,x_{i,l})+\varepsilon_{i}(u), \quad i\in G_{k}, \quad k=1,...,K.
\end{equation}

To estimate the group-specific functions $ m_{k|K,0}(u)$ and bivariate additive component functions $ g_{l}(u,x_{l})$, we apply the backfitted local linear algorithm proposed by \cite{f93}. Define
\begin{align*}
f_{i,0}^{c}(u)=&f_{i}(u)-\sum_{l=1}^{p}g_{l}(u,x_{i,l})= m_{k|K,0}(u)+\varepsilon_{i}(u), \\
f_{i,l}^{c}(u,x_{i,l})=&f_{i}(u)-m_{k|K,0}(u)-\sum_{j\neq l}^{p}g_{j}(u,x_{i,j})= g_{l}(u,x_{i,l})+\varepsilon_{i}(u), \quad i\in G_k.
\end{align*}

By plugging in the initial estimation $ \hat{g}_{l}(u,x_{i,l}) $, we obtain the estimation of $ f^{c}_{i,0}(u) $ as
\begin{equation*}
\hat{f}_{i,0}^{c}(u)=f_{i}(u)-\sum_{l=1}^{p}\hat{g}_{l}(u,x_{i,l}).
\end{equation*}

For each given $ u_{t}\in [0,1] $, $m_{k|K,0}(u_{t}) $ can be approximated by the first-order Talyor expansion
\begin{align*}
m_{k|K,0}(u_{t})\approx m_{k|K,0}(u)+h_{0}\dfrac{\partial m_{k|K,0}(u)}{\partial u}\dfrac{u_{t}-u}{h_{0}}\triangleq a_{k,0}+b_{k,0}\dfrac{u_{t}-u}{h_{0}}.
\end{align*}

Define the weighted squared function as
\begin{equation*}
Q_{0}(a_{k,0},b_{k,0})=\sum_{i\in \hat{G}_{k}}\sum_{t=1}^{T}\biggl[\hat{f}_{i,0}^{c}(u_{t})-a_{k,0}-\dfrac{u_{t}-u}{h_{0}}b_{k,0}\biggl]^2\mathcal{K}(\dfrac{u_{t}-u}{h_{0}}),\qquad k=1,...,K,
\end{equation*}
where $ \mathcal{K}(\cdot) $ is a nonnegative kernel function with bandwidth $ h_0 $ and $ \hat{G}_{k} $ denotes the estimated membership of group $ k $. Then by minimizing $Q_{0}(a_{k,0},b_{k,0})$, the estimation of $ m_{k|K,0}(u) $ as $ \hat{m}_{k|K,0}(u)=\hat{a}_{k,0} $ is derived as
\begin{equation}\label{eq:local-m}
\hat{m}_{k|K,0}(u)=\sum_{i\in \hat{G}_{k}}\sum_{t=1}^{T}w_{0,t}\hat{f}^{c}_{i,0}(u_{t})/\sum_{t=1}^{T}w_{0,t}, \qquad k=1,...,K,
\end{equation}
where $ w_{0,t}=\mathcal{K}(\dfrac{u_{t}-u}{h_{0}})(c_{0,2}-\dfrac{u_{t}-u}{h_{0}}c_{0,1}) $ and $ c_{0,j}=\sum_{t=1}^{T}\mathcal{K}(\dfrac{u_{t}-u}{h_{0}})(\dfrac{u_{t}-u}{h_{0}})^{j} $ for $ j=1,2 $.

With the estimation of group-specific functions, we can have the estimation of $ f_{i,l}^{c}(u,x_{i,l}) $
\begin{equation*}
\hat{f}_{i,l}^{c}(u,x_{i,l})=f_{i}(u)-\sum_{k=1}^K\hat{m}_{k|K,0}(u)I(i\in\hat{G}_{k})-\sum_{j\neq l}^{p}\hat{g}_{j}(u,x_{i,j}),
\end{equation*}
where $I(\cdot)$ is the indicator function.

For each given $ u\in [0,1] $ and $ X_{i,l}\in [0,1] $, we approximate $ g_{l}(u,X_{i,l}) $ by
\begin{equation*}
g_{l}(u,X_{i,l})\approx g_{l}(u,x_{l})+h_{l,x}\dfrac{\partial g_{l}(u,x_{l})}{\partial x_{l}}\dfrac{X_{i,l}-x_{l}}{h_{l,x}}\triangleq a_{l}+b_{l}\dfrac{X_{i,l}-x_{l}}{h_{l,x}}.
\end{equation*}

Then the pointwise estimator of $ g_{l}(u,x_{l}) $ as $ \tilde{g}_{l}(u,x_{l})=\hat{a}_{l}$ can be obtained by minimizing
\begin{equation*}
\sum_{i=1}^{n}\biggl[\hat{f}_{i,l}^{c}(u,X_{i,l})-a_{l}-\dfrac{X_{i,l}-x_{l}}{h_{l,x}}b_{l}\biggl]^2\mathcal{K}(\dfrac{X_{i,l}-x_{l}}{h_{l,x}}).
\end{equation*}

Since the pointwise estimation of $ g_{l}(u,x_{l}) $ for each $ u\in [0,1]$ may not be smooth due to the dependence on the estimation of density responses $ f_{i} $, then an additional local smoothing step is implemented in the direction of $ u $ to rectify the smoothness. To do so, for each $ u\in[0,1] $, we define the following function
\begin{equation*}
Q_{l}(a_{l},b_{l})=\sum_{i=1}^{n}\int_{0}^{1}\biggl[\hat{f}_{i,l}^{c}(v,X_{i,l})-a_{l}-\dfrac{X_{i,l}-x_{l}}{h_{l,x}}b_{l}\biggl]^2\mathcal{K}
(\dfrac{X_{i,l}-x_{l}}{h_{l,x}})\mathcal{W}(\dfrac{u-v}{h_{l,u}})dv,
\end{equation*}
where $ \mathcal{W}(\cdot) $ is a nonnegative kernel function with bandwidth $ h_{l,u} $. By minimizing $Q_{l}(a_{l},b_{l})$, the refined estimator of $ g_{l}(u,x_{l}) $ can be obtained as
\begin{equation}\label{eq:local-g}
\tilde{g}_{l}(u,x_{l})=\sum_{i=1}^{n}w_{l,i}\int_{0}^{1}\hat{f}^{c}_{i,l}(v,X_{i,l})\mathcal{W}(\dfrac{u-v}{h_{l,u}})dv\biggl/\int_{0}^{1}\mathcal{W}(\dfrac{u-v}{h_{l,u}})dv\sum_{i=1}^{n}w_{l,i}, \quad l=1,...,p,
\end{equation}
where $ w_{l,i}=\mathcal{K}(\dfrac{X_{i,l}-x_{l}}{h_{l,x}})(c_{l,2}-\dfrac{X_{i,l}-x_{l}}{h_{l,x}}c_{l,1}) $ and $ c_{l,j}=\sum_{i=1}^{n}\mathcal{K}(\dfrac{X_{i,l}-x_{l}}{h_{l,x}})(\dfrac{X_{i,l}-x_{l}}{h_{l,x}})^{j} $ for $ j=1,2 $.

%%%%%%%%%%%%%%%%%%%%%%%%%%%%%%%%%%%%%%%%%%%%%%%%%%%%%%%%%%%%%%%%%%%%%%%%%%%%%%%%%%%%%%%%%%%%%%%%%%%%%%%%%%%%%%%%%%
\subsection{Selection of Number of Groups}\label{sec:information}

The identification and estimation of the latent group structures discussed above is based on the prior that the number of groups $ K $ is known, but it is not true for practical problem. A major challenge in clustering is to estimate the group when the number is unknown. Following \cite{c19}, the information criterion is applied for the selection of group number.  

Denote
\begin{equation*}
 IC(\dot{K})=\log V^{2}_{n}(\dot{K})+\dot{K}\cdot \rho,
\end{equation*}
where
\begin{equation*}
V^{2}_{n}(\dot{K})=\frac{1}{nT}\sum_{k=1}^{\dot{K}}\sum_{i\in \hat{G}_{k}}\sum_{t=1}^{T}\biggl|\biggl|\hat{f}_{i,0}^{c}(u_{t})-\hat{m}_{k|\dot{K},0}(u_{t})\biggl|\biggl|_{2}^{2},
\end{equation*}
and $ \rho $ is a tuning parameter whose value may rely on $ n $.

The number of latent groups is estimated by minimizing the criterion $ IC(\dot{K}) $, i.e.,
\begin{equation*}
\hat{K}=\arg \min_{1\leq \dot{K} \leq \tilde{K}} IC(\dot{K}),	
\end{equation*}
where $ \tilde{K} $ is pre-specified as the maximal number of groups.

In the simulation studies, we take two choices of $ \rho $ into account. They are
\begin{equation*}
\rho_{1}=\dfrac{\log(n_{\dot{K}}Th)}{n_{\dot{K}}Th}\,\quad \text{and}\, \quad
\rho_{2}=\dfrac{2}{n_{\dot{K}}Th},
\end{equation*}
where $ n_{\dot{K}}=\min\{|\hat{G}_{\dot{K}}|,k=1,...,\dot{K}\} $ and $ |\hat{G}_{k}| $ denotes the cardinality of the group $ \hat{G}_{k} $. They correspond to two information criterions, the generalized Bayesian information criterion (GBIC) with $ \rho=\rho_{1} $ and the generalized Akaike information criterion(GAIC) with $ \rho=\rho_{2} $. Other information criterions can be found in \cite{clwz19} and \cite{c19}.

%%%%%%%%%%%%%%%%%%%%%%%%%%%%%%%%%%%%%%%%%%%%%%%%%%%%%%%%%%%%%%%%%%%%%%%%%%%%%%%%%%%%%%%%%%%%%%%%%%%%%%%%%%%%%%%%%%
\subsection{Selection of Bandwidth}\label{sec:bandwidth}

In this article, the leave-one-out cross-validation method is implemented for the bandwidth selection. Let $\boldsymbol{h}=(h_{0}, h_{l,u}, h_{l,x})^{\tau}$, we select the bandwidth $ \boldsymbol{h} $ given the number of latent group $ K $ by minimizing the following mean squared error
\begin{equation*}
CV(\boldsymbol{h})=\dfrac{1}{nT}\sum_{i=1}^{n}\sum_{t=1}^{T}[f_{i}(u_t)-\sum_{k=1}^{K}\hat{m}^{(-t,\boldsymbol{h})}_{k|K,0}(u_{t})I(i\in \hat{G}_{k})-\sum_{l=1}^{p}g^{(-t,\boldsymbol{h})}_{l}(u_{t},X_{i,l})]^{2}, 
\end{equation*}
where $ \hat{G}_{k} $ is the estimated group membership. For each $ t=1,...,T $, $ \hat{m}^{(-t,\boldsymbol{h})}_{k|K,0}(u_{t}) $ and $ g^{(-t,\boldsymbol{h})}_{l}(u_{t},X_{i,l}) $ are the estimations with bandwidth $ \boldsymbol{h} $ of $ m_{k|K,0}(u_{t}) $ and $ g_{l}(u_{t},X_{i,l}) $ obtained by using observations except the $ t $-th observation respectively.

%%%%%%%%%%%%%%%%%%%%%%%%%%%%%%%%%%%%%%%%%%%%%%%%%%%%%%%%%%%%%%%%%%%%%%%%%%%%%%%%%%%%%%%%%%%%%%%%%%%%%%%%%%%%%%%%%%
\section{Theoretical Results}

Throughout this paper, for any fixed interval $ [a,b] $, we denote the space of $ l$-th order smooth function as $ C^{(l)}[a,b]=\{g|g^{(l)}\in c[a,b]\} $, and the class of Lipschitz continuous functions for some fixed constant $ C>0 $ as $ Lip([a,b],C)=\{g||g(x)-g(x^{'})|\leq |x-x^{'}|, \forall x, x^{'}\in [a,b]\}. $ Meanwhile, let $ S_{l} $ and $ S_{x} $ denote the support of $ x_{l} $ and $ \boldsymbol{x} $, respectively. Obviously, $ S_{x}=\prod_{l=1}^{p}S_{l} $. The necessary assumptions for the asymptotic results are listed as follows.

(A1) For any $ z\in \mathcal{F} $, $ z $ is differentiable, and there exists a constant $ M>1 $, such that $ ||z||_{\infty}, ||1/z||_{\infty},||z^{'}||_{\infty}$ are all bounded by $ M $.

(A2) (a) The kernel density $ \mathcal{K} $ is Lipschitz-continuous, bounded and symmetric about 0. Furthermore, $ \mathcal{K} \in Lip([-1,1],L_{k}) $ for some constant $ L_{k}>0 $. (b) The kernel density $ \mathcal{K} $ satisfies the conditions that $ \int_{0}^{1}\mathcal{K}(u)du>0 $, $ \int_{\mathcal{R}}^{}|u|\mathcal{K}(u)du $, $ \int_{\mathcal{R}}^{}\mathcal{K}^{2}(u)du $ and $ \int_{\mathcal{R}}^{}|u|\mathcal{K}^{2}(u)du $ are finite. The kernel density $ \mathcal{W} $ also satisfies the above assumptions.

(A3) (a) For each $ 1\leq i\leq n $, the process $ \{\boldsymbol{x_{i}},\varepsilon_{i}(u)\} $ is stationary and $ \alpha$-mixing dependent for $ u \in [0,1] $, with the mixing coefficient decaying to zero at a geometric rate, i.e., there exists constants $ A< \infty $ and $ \beta>4 $, such that $ \alpha(k)\leq Ak^{-\beta} $ for all $ k \geq 1 $. (b) The covariate variables $ x_{i,l} $, $ 1\leq l\leq p $, and the errors $ \varepsilon_{i}(u) $ satisfy the following moment conditions that for some $ s>2 $,
\begin{equation*}
\max_{1 \leq i \leq n}\max_{1 \leq l \leq p}E(|x_{i,l}|^{2s})<\infty, \quad \max_{1 \leq i \leq n}E(|\varepsilon_{i}(u)|^{2s})<\infty.
\end{equation*}
For each $ i=1,...,n $, the covariance function $ Cov(\varepsilon_{i}(s),\varepsilon_{i}
(t)|\boldsymbol{x_{i}})=\boldsymbol{\Sigma}_{i}(s,t)$ has finite non-decreasing eigenvalues $ \lambda_{1}\leq...\leq \lambda_{max} $, satisfying $ \sum_{j}\lambda_{j}<\infty $.

(A4) The latent group functions $ m_{k|K,0}(\cdot) $, $ 1 \leq k \leq K $, have continuous second order derivatives on the support interval, namely, $ m_{k|K,0}(\cdot) \in C^{(2)}[0,1] $, and $ m^{'}_{k|K,0}(\cdot) \in Lip([0,1],L_{0})  $ for some constant $ L_{0}>0 $. Meanwhile, the additive component functions $ g_{l}(u,x_{l}), 1 \leq l \leq p $ are continuous functions on $[0,1]\times[a_{l},b_{l}]$ and twice continuously partial differentiable with respective to $ u $ and $ x_{l} $, where $ [a_{l},b_{l}] $ is a compact subset of $ S_{l} $.

(A5) (a) The density function of covariate $ \boldsymbol{x} $, $ f(\boldsymbol{x}) $,  is continuous and bounded, with continuous derivatives of marginal densities $ f_{l,u}(x_{l}) $ at each $ u\in[0,1] $.
(b) For $ u\in [0,1] $ and $ \boldsymbol{x} \in S_{x} $, the joint density of $ u $ and $\boldsymbol{x} $, $ f(u,\boldsymbol{x}) $, as well as the joint density of $ u $ and $ x_{l} $, $ f_{l}(u,x_{l}) $, are continuous and partially differentiable with respect to $ u $ and $ \boldsymbol{x} $, with continuous second order partial derivatives.

(A6) (a) Let $ \delta=\min_{1\leq k\neq l \leq K}\min_{i \in G_{k}, j\in G_{l}}d_{i,j}$, where $ d_{i,j} $ is the $(i,j)$ element of distance matrix discussed before, then $ h^{2}+(Th)^{-1/2}=o_{p}(\delta) $. (b) There exists a positive constant $ \xi $, with $ 0<\xi <1 $, such that$ \min_{1\leq k\leq K}|G_{k}|\geq \xi \cdot n $.

(A7) $ N_{0}\sim T^{1/5} $, $ N_{l}\sim T^{1/6} $, $h_{0}\sim (nT)^{-1/5}$, $ h_{l,u}, h_{l,x}\sim n^{-1/5}$, $ 1\leq l\leq p $, as $ n,T \rightarrow \infty $.

\begin{rem} Assumption (A1) is basic and essential to derive the consistency of densities after transformation. The conditions in (A2) on the kernel function $ \mathcal{K}(\cdot) $ are mild and can be satisfied by commonly used kernel functions such as uniform and Epanechnikov kernel. Assumption (A3) (a) relaxes the dependence of error process and covariates spaces to the $ \alpha$-mixing dependence which is one of the weakest dependence conditions. The moment conditions in (A3) (b) is crucial to derive the uniform convergence and other asymptotic properties based on the kernel function. The smoothness conditions of component functions in (A4) and (A5) are greatly relaxed. Assumption (A6) (a) indicates that $ \delta $ can converge to zero at an appropriate rate, and (b) are useful in proving the consistency of estimated group number $ \hat{K} $ via information criterion proposed before. (A7) are common conditions applied in kernel smoothing to satisfy the optimal convergence rates.
\end{rem}
%%%%%%%%%%%%%%%%%%%%%%%%%%%%%%%%%%%%%%%%%%%%%%%%%%%%%%

We first derive the uniform consistency of initial estimations of $ g_{i,0}(u) $ and $ g_{l}(u,x_{i,l}) $ which is presented by
Theorem \ref{thm:g_initial}.

\begin{thm}\label{thm:g_initial}
Assume that (A1)$\sim$(A4) and (A7) hold, $ \hat{g}_{i,0}(u) $ and $ \hat{g}_{l}(u,x_{i,l}) $ are the initial estimations of $ g_{i,0}(u) $ and $ g_{l}(u,x_{i,l}) $ respectively, which are defined by (\ref{eq:initial}), $ i=1,...,n,$ ,$ l=1,...,p $. Then as $ T \rightarrow \infty $ and $ n \rightarrow \infty $, it holds that
\begin{align*}
&(i) \quad \sup_{u}|\hat{g}_{i,0}(u)-g_{i,0}(u)|=O_{p}(T^{-2/5}\log{T}+h^{2}+(Th)^{-1/2});\\
&(ii) \quad \sup_{u,x_{i,l}}|\hat{g}_{l}(u,x_{i,l})-g_{l}(u,x_{i,l})|=O_{p}(T^{-1/3}\log{T}+h^{2}+(Th)^{-1/2}).
\end{align*}
\end{thm}

Theorem \ref{thm:clust} and \ref{thm:kconsist} claim that the number of groups $K$ and the membership of group structures $\{ G_{1},...,G_K\}$ can be correctly identified with probability 1.

%%%%%%%%%%%%%%%%%%%%%%%%%%%%%%%%%%%%%%%%%%%%%%%%%%%%%

\begin{thm}\label{thm:clust}
	Assume that (A1)$\sim$(A5) hold and the number of latent groups $ K $ is known. Denote $ \{G_{1},...,G_{K}\}$ the true group structure of $ g_{i,0}(\cdot) $, $ i=1,...,n $, and $ \{\hat{G}_{1},...,\hat{G}_{K}\} $  the corresponding estimation. Then as $ T\rightarrow \infty $, it holds that
	\begin{equation*}
	P(\{\hat{G}_{1},...,\hat{G}_{K}\}=\{G_{1},...,G_{K}\})\rightarrow 1.
	\end{equation*}
\end{thm}

\begin{thm}\label{thm:kconsist}
Suppose that (A1)$ \sim $ (A7) hold. If $ K$ is the number of latent groups, and $\hat{K} $ is the estimation of $ K $ via information criterion, then
\begin{equation*}
P(\hat{K}=K)\rightarrow1, \text{as $ T\rightarrow \infty $}.
\end{equation*}

\end{thm}

%%%%%%%%%%%%%%%%%%%%%%%%%%%%%%%%%%%%%%%%%%%%%%%%%%%%

\begin{thm}\label{thm:local}
 Assume that (A1)$ \sim $(A7) hold and the number of latent groups $ K $ is known. Then, as $ n \rightarrow\infty $ and $ T \rightarrow\infty $ , it holds that \\
(i)
\begin{equation*}
\sup_{u}|\hat{m}_{k|K,0}(u)-m_{k|K,0}(u)|=O_{p}\biggl((nT)^{-2/5}(\log nT)^{1/2}+h^{2}+(Th)^{-1/2}\biggl);
\end{equation*}
(ii)
\begin{equation*}
\sup_{u,x_{l}}|\tilde{g}_{l}(u,x_{l})-g_{l}(u,x_{l})|=O_{p}\biggl(n^{-2/5}(\log n)^{1/2}+h^{2}+(Th)^{-1/2}\biggl).
\end{equation*}
\end{thm}

Theorem \ref{thm:local} characterizes the uniform convergence of the oracle post-clustering estimation of group-specific functions and additive components. Subsequently, to establish the asymptotical normality of $\hat{m}_{k|K,0}(u)$ and $\tilde{g}_{l}(u,x_{l}) $, we define
$$ \mu_{j}=\int \mathcal{K}(x)x^{j}dx<\infty ,\quad \iota_{j}=\int \mathcal{W}(x)x^{j}du<\infty , \quad  \nu_{j}=\int \mathcal{K}^{2}(x)x^{j}dx<\infty. $$

\begin{thm}\label{thm:asymptotic}
Assume that (A1)$ \sim $(A7) hold. $ \hat{K} $ is the estimation of the number of latent groups, then  Then,  as $ T \gg n \rightarrow\infty $, it holds for all $ u\in (0,1) $ and $ x_{l}\in [0,1] $, that\\
(i)
\begin{equation*}
\sqrt{nTh_{0}}(\hat{m}_{k|\hat{K},0}(u)-m_{k|K,0}(u)-B_{k,0}(u))\xrightarrow{D}N(0,V_{0}(u)),
\end{equation*}
where $ B_{k,0}(u)=\dfrac{\mu_{2}h_{0}^{2}}{2}m^{''}_{k|K,0}(u) $, $ V_{0}(u)=\sigma^{2}(u)\nu_{0}^{2} $.

(ii)
\begin{equation*}
\sqrt{nh_{l,x}}(\tilde{g}_{l}(u,x_{l})-g_{l}(u,x_{l})-B_{l}(u,x_{l}))\xrightarrow{D}N(0,V_{l}(u,x_{l})),
\end{equation*}
where $B_{l}(u,x_{l})=\dfrac{1}{2}(\mu_{2}h_{l,u}^{2}\dfrac{\partial^{2}g(u,x_{l})}{\partial^{2}u}+\iota_{2}h_{l,x}^{2}\dfrac{\partial^{2}g(u,x_{l})}{\partial^{2}x_{l}}) $, $ V_{l}(u,x_{l})=\sigma^{2}(u)\nu_{0}/f_{l,u}(x_{l})$.
\end{thm}

%%%%%%%%%%%%%%%%%%%%%%%%%%%%%%%%%%%%%%%%%%%%%%%%%%%%%%%%%%%%%%%%%%%%%%%%%%%%%%%%%%%%%%%%%%%%%%%%%%%%%%%%%%%%%%%%%%
\section{Numerical Study}

In this section, we conduct the simulation study to demonstrate the performance of the proposed identification and estimation procedure for the model \eqref{eq:model}. In this simulation, we consider the case that the group number $K=3$ and covariates number $p=2$. The regression model \eqref{eq:model} with the latent group structures \eqref{eq:latentstructure} is
\begin{equation*}
f_{i}(u)=\mu_i(u|\boldsymbol{X}_{i})+\varepsilon_i(u)=g_{i,0}(u)+g_{1}(u,x_{i,1})+g_{2}(u,x_{i,2})+\varepsilon_{i}(u), \quad 1\leq i\leq n,
\end{equation*}
where $\mu_i(u|\boldsymbol{X}=\boldsymbol{x})=g_{i,0}(u)+g_{1}(u,x_{i,1})+g_{2}(u,x_{i,2})$, with the latent group structures
\begin{equation*}
g_{i,0}(u)=\left\lbrace
\begin{aligned}
& m_{1,0}(u)=\sqrt{2}\sin(2\pi u),  & i\in G_{1},\\
& m_{2,0}(u)=\sqrt{2}\cos(2\pi u),  & i\in G_{2},\\
& m_{3,0}(u)=6[2u-6u^{2}+4u^{3}+0.05], & i\in G_{3},
\end{aligned}
\right.
\end{equation*}
and the additive component functions
\begin{equation*}
g_{1}(u,x_{1})=\sin(2\pi u)(2x_{1}-1), \quad
g_{2}(u,x_{2})=\sin(2\pi u)\sin(2\pi x_{2}),
\end{equation*}
for $ u,x_{1},x_{2}\in[0,1] $. The groups are defined as $ G_{1}=\{1,2,...,n_{1}\} $, $ G_{2}=\{n_{1}+1,n_{1}+2,...,n_{1}+n_{2}\} $, $ G_{3}=\{n_{1}+n_{2}+1,n_{1}+n_{2}+2,...,n_{1}+n_{2}+n_{3}\}$, and the cardinalities of each group is set to be $ n_{1}=0.3n, n_{2}=0.3n, n_{3}=0.4n $.

The covariates $x_{i,1},x_{i,2}$ are generated by $ \boldsymbol{x_{i}}=(x_{i,1},x_{i,2})^{\tau}=(\Phi(v_{i,1}),\Phi(v_{i,2}))^{\tau} $, $ 1\leq i\leq n $, where $ \Phi $ is the CDF of standard normal distribution, $ \boldsymbol{v_{i}}=(v_{i,1},v_{i,2})^{\tau}\stackrel{iid}{\sim} N_{2}(\boldsymbol{0},\Sigma) $ are bivariate normal random vectors with mean zero and covariance matrix
$ \Sigma=\begin{pmatrix}
1   &  0.5 \\
0.5 &  1 
\end{pmatrix} $. The random error $ \varepsilon(u)=\epsilon_{1}sin(\pi u)+\epsilon_{2}sin(2\pi u) $, where $ \epsilon_{1} \sim N(0,0.1^{2})$, $ \epsilon_{2} \sim N(0,0.05^{2})$, and $\epsilon_{1}$ are independent of $ \epsilon_{2}$.

We note that the conditional mean functions $\mu(u|\boldsymbol{X}=\boldsymbol{x})$ are the log-quantile transformation of the random density $z(u|\boldsymbol{x})$. More specifically, the inverse of log-quantile transformation is $\Psi^{-1}(\mu(u|\boldsymbol{x}))=\theta(\boldsymbol{x})\exp\{-\mu(F(u|\boldsymbol{x}),\boldsymbol{x})\} $, where $ \theta(\boldsymbol{x})=\int_{0}^{1}\exp\{\mu(v,\boldsymbol{x})\}dv $. Thus the conditional distribution function $ F(\cdot|\boldsymbol{x}) $ and quantile function $ Q(\cdot|\boldsymbol{x}) $ satisfy
\begin{equation*}
Q(u|\boldsymbol{x})=F^{-1}(u|\boldsymbol{x})=\theta(\boldsymbol{x})^{-1}\int_{0}^{u}\exp\{\mu(v,\boldsymbol{x})\}dv.
\end{equation*}
where $ \theta(\boldsymbol{x})=\int_{0}^{1}\exp\{\mu(v,\boldsymbol{x})\}dv $.

To generate the observations of response, for each $ 1\leq i \leq n $, let $ u_{i,1},...,u_{i,T_{i}}\sim \text{Uniform}(0,1) $, which are independent of $ \boldsymbol{X_{i}} $, the observations of $f_i$ at time points $ u_{i,1},...,u_{i,T_{i}}$ are $ \boldsymbol{{y_{i}}}=\{y_{i,j}=Q(u_{i,j}|\boldsymbol{X_{i}}):1\leq j\leq T_{i}\} $, thus $ Y_{i,1},...,Y_{i,T_{i}}\stackrel{iid}{\sim} z_{i}\equiv\Psi^{-1}(g(\cdot,\boldsymbol{X_{i}})+\varepsilon_{i}(\cdot)) $, where $ z_{i} $ are the random response densities. Without loss of generality, we assume $ T_{i} =T $. We set the sample size $ n=50,100$, the number of observations from each random density is $ T=50,100 $. For each setting, the simulation repeated 200 times.

\begin{figure}[!ht]\centering
	\includegraphics[width=15cm]{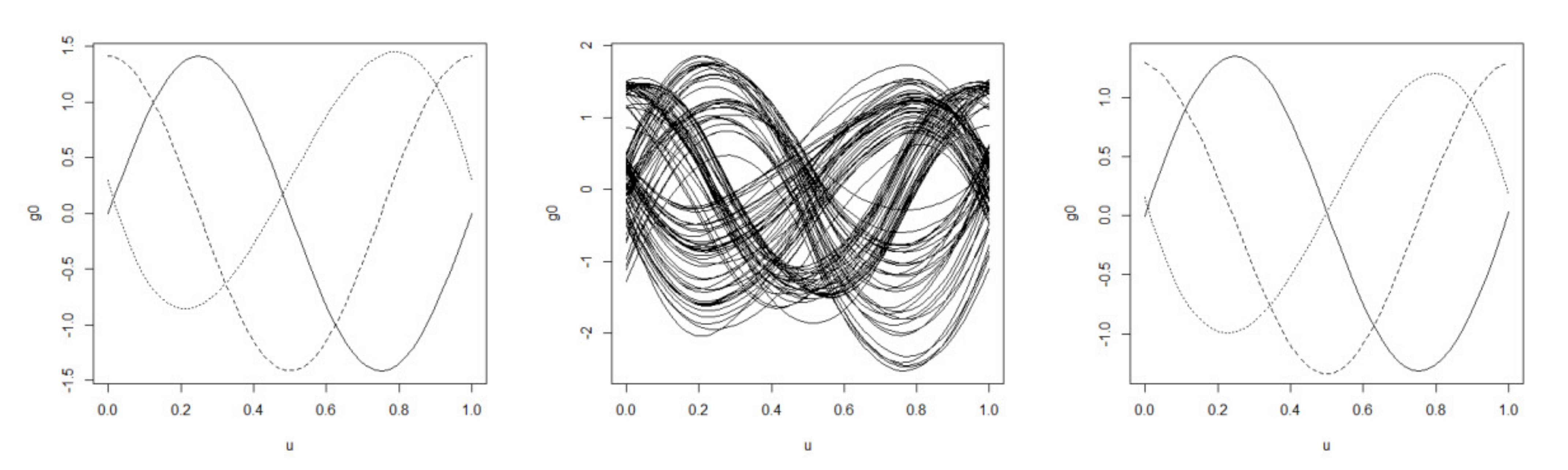}
	\caption{The average estimation of latent group structure term $ g_{i,0}(\cdot) $ obtained from 200 Monte Carlo runs with sample sizes $ n=100 $ and $ T=100 $. Left panel: true densities, middle panel: pre-clustering estimations, right panel: post-clustering estimations.  } \label{fg:g0}
\end{figure}

Figures \ref{fg:g0} and A.1 depict the average performance of pre- and post- clustering estimations for group structure terms $g_{i,0}(u)$ and bivariate additive components $g_{l}(u,x_{l})$, respectively, through 200 Monte Carlo runs with sample sizes $ n=100 $ and the number of time points $ T=100 $. In each figure, the true functions, pre-clustering and post-clustering estimations are presented from the left to right panels.

From Figure \ref{fg:g0} we can see that although the pre-clustering estimations capture the basic shapes of true densities, but there exists certain differences between the true curves and pre-clustering estimations. The minimizer and maximizer for each estimated curve are close to the true points but the extreme values between estimation and true curve do exist. On the other hand, post-clustering estimations more accurately depict the true density curves since not only they catch the shape of curves but also the minimizer, maximizer and extreme values for each estimated curve are almost same as the true density curve.
Similar patterns can be found from  the estimations of bivariate additive functions in Figure A.1, meaning the efficiency of proposed identification procedure under this setting.

Denote $ C=\{G_{1},...,G_{K}\} $ the true clusters and $ \hat{C}=\{\hat{G}_{1},...,\hat{G}_{\hat{K}}\} $ the estimated clusters. To evaluate the performance of the clustering algorithm, we take two measures into account. One is the traditional measure for clustering, the purity, which is defined as
\begin{equation*}
Purity(\hat{C})=\frac{1}{n}\sum_{k=1}^{\hat{K}}\max_{1 \leq j \leq K}|\hat{G}_{k}\cap G_{j}|.
\end{equation*}
The other one is the normalized mutual information (NMI) which is a common measure for the similarity between clusterings \citep{kfw15}. Here, we define NMI between $\hat{C}$ and the true clusters $C$, i.e.,
\begin{equation*}
NMI(\hat{C},C)=\dfrac{I(\hat{C},C)}{(H(\hat{C})+H(C))/2},
\end{equation*}
where $ I(\hat{C},C) $ is the mutual information between $ \hat{C} $ and $ C $ which is defined as
\begin{equation*}
I(\hat{C},C)=\sum_{k=1}^{\hat{K}}\sum_{j=1}^{K}\bigg(\dfrac{|\hat{G}_{k}\cap G_{j}|}{n}\bigg)\log_{2}\bigg(\dfrac{n|\hat{G}_{k}\cap G_{j}|}{|\hat{G}_{k}||G_{j}|}\bigg),
\end{equation*}
\begin{equation*}
H(\hat{C})=-\sum_{k=1}^{\hat{K}}\frac{|\hat{G}_{k}|}{n}\log_{2}\bigg(\frac{|\hat{G}_{k}|}{n}\bigg),
\end{equation*}
is the entropy of $ \hat{C} $ and $ H(C) $ is defined analogously. Since both Purity and NMI do not depend on the ordering of clusters, they are proper measures for the efficiency of the clustering algorithm. It is obvious that the closer the both values are to 1, the more efficient the algorithm is, namely, the closer the estimated clusters are to the true clusters.

To evaluate the efficiency of estimation procedure, we compare the performance of three estimators. one is $(\hat{g}_{i,0}(u), \hat{g}_l(u,x_l))$, the pre-clustering estimator obtained without considering the group structures, second is $(\hat{m}_{k|K,0}(u), \tilde{g}_l(u,x_l))$ with given $K$, the oracle estimator obtained by giving the number of true groups, and the last one is $(\hat{m}_{k|\hat{K},0}(u), \tilde{g}_l(u,x_l))$, post-clustering obtained from data in each estimated group. The root mean squared errors (RMSEs) are used to examine the performance of the estimations. For instance, the RMSE of the pre-clustering estimator are defined as
\begin{equation*}
RMSE(\hat{g}_{0})=\frac{1}{n}\sum_{i=1}^{n}\biggl\{\frac{1}{T}\sum_{t=1}^{T}||\hat{g}_{i,0}(u_{t})-g_{i,0}(u_{t})||^{2}_{2}\biggl\}^{\frac{1}{2}},
\end{equation*}
\begin{equation*}
RMSE(\hat{g}_{l})=\frac{1}{n}\sum_{i=1}^{n}\biggl\{\frac{1}{T}\sum_{t=1}^{T}||\hat{g}_{l}(u_{t},x_{i,l})-g_{l}(u_{t},x_{i,l})||^{2}_{2}\biggl\}^{\frac{1}{2}}.
\end{equation*}
The RMSEs of the other estimators are defined analogously.

\begin{table}[!ht]
 \centering
 \setlength{\tabcolsep}{4.7mm}{
 \begin{tabular}{@{\extracolsep{\fill}}*{12}{c}}\hline \hline 	
	\multicolumn{12}{c}{Account of the estimated value of $K$} \\\hline
	\multicolumn{2}{c}{Sample Size} & \multicolumn{5}{c}{GAIC} & \multicolumn{5}{c}{GBIC} \\\hline
	$ n $ & $ T$ &  $ 1 $ &  $2 $ & $3 $ &  $4$ &  $5 $ & $ 1 $ &  $2 $ & $3 $ &  $4$ &  $5 $\\\hline
	50&	50&	0&	12&	183&	5&	0&	0&	6&	185&	9&	0\\
	&  100&	0&	1&	199&	0&	0&	0&	1&	199&	0&	0\\
   100& 50&	0&	10&	187&	3&	0&	0&	5&	188&	7&	0\\
	&  100&	0&	0&	200&	0&	0&	0&	0&	200&	0&	0\\\hline
 \end{tabular}}
\setlength{\tabcolsep}{3mm}{
 \begin{tabular}{@{\extracolsep{\fill}}*{6}{c}}\hline
	\multicolumn{6}{c}{Average(standard deviation) of NMIs and Purities} \\\hline
	\multicolumn{2}{c}{Sample Size} & \multicolumn{2}{c}{GAIC} & \multicolumn{2}{c}{GBIC} \\\hline		
	$ n $ & $ T $ &  $ NMI $ &  $Purity $ & $NMI $ &  $Purity$ \\\hline
    50&	50&	0.8351(0.0621)&	0.9236(0.0582)&	0.8273(0.0653)&	0.9275(0.0619)\\
	  &100&	0.9562(0.0584)&	0.9831(0.0469)&	0.9562(0.0584)&	0.9831(0.0469)\\
   100&	50&	0.8617(0.0531)&	0.9527(0.0426)&	0.8561(0.0526)&	0.9513(0.0485)\\
	  &100&	0.9835(0.0392)&	0.9962(0.0273)&	0.9835(0.0392)&	0.9962(0.0273)\\\hline
 \end{tabular}	}
\caption{The account of the estimated value of $K$, average and standard deviation of NMIs and Purities.} \label{tab:results1}
\end{table}

\begin{table}[!ht]
\centering
\setlength{\tabcolsep}{3mm}{
 \begin{tabular}{@{\extracolsep{\fill}}*{6}{c}}\hline
 	\multicolumn{6}{c}{Averages(standard deviations) of RMSEs for the estimation of $ g_{0}(u) $} \\\hline
 	\multicolumn{2}{c}{Sample Size} & \multicolumn{4}{c}{$ g_{0}(u) $} \\\hline
 	\multicolumn{2}{c}{}& \multicolumn{1}{c}{Oracle} & \multicolumn{1}{c}{Pre-clustering} & \multicolumn{2}{c}{Post-clustering} \\\cline{3-6}		
 	$ n $ & $T $ &  $  $ &  $ $ & $ GAIC $ &  $GBIC$ \\\hline
    50&	50&	0.2869(0.0507)&	0.4725(0.0531)&	0.3215(0.0628)&	0.3167(0.0615)\\
	  &100&	0.2381(0.0319)&	0.4128(0.0342)&	0.2463(0.0437)&	0.2463(0.0437)\\
   100&	50&	0.2537(0.0431)&	0.4531(0.0462)&	0.2618(0.0535)&	0.2637(0.0526)\\
	  &100&	0.2125(0.0253)&	0.3962(0.0275)&	0.2179(0.0312)&	0.2179(0.0312)\\\hline		  	
	\end{tabular}}

 \caption{The average and standard deviation of RMSEs of the latent group structures} \label{tab:results2-g0}
\end{table}	

Table \ref{tab:results1}, \ref{tab:results2-g0} and A.1 present the results under different settings, which include account of the estimated value of $K$, average and standard deviation of NMIs, Purities and RMSEs of the estimations. Firstly, about the performance of the clustering algorithm, we can see that the overall performances of GAIC and GBIC are quite similar. Besides, under either GAIC or GBIC, the account that $\hat{K}$ equals the true value $K$, NMI, and purity rises with the increase of $n$ or $T$. When $T=100$ and $n=100$, the value $K$ is one hundred percent truly estimated, and the NMI and purity is close to 1. Secondly, about the performance of the estimators, it is clear from Table \ref{tab:results2-g0} and A.1
that the mean and standard deviation of RMSE for all estimators decrease with the increase of sample size $ n $ and number of time points $ T $. Under all scenarios, the oracle and post-clustering estimators outperform the pre-clustering estimators. The RMSEs of post-clustering estimators are quite similar as oracle estimators, and they are getting closer along with the increase of $n$ and $T$. Especially when $n=100,T=100$, they are very close to each other. Moreover, the performance of post-clustering estimators under GAIC and GBIC criterions are almost same.

%%%%%%%%%%%%%%%%%%%%%%%%%%%%%%%%%%%%%%%%%%%%%%%%%%%%%%%%%%%%%%%%%%%%%%%%%%%%%%%%%%%%%%%%%%%%%%%%%%%%%%%%%%%%%%%%%%
\section{Real Data Analysis}

In this section, we apply the proposed methodology to two social studies.
%%%%%%%%%%%%%%%%%%%%%%%%%%%%%%%%%%%%%%%%%%%%%%%%%%%%%%%%%%%%%%%%%%%%%%%%%%%%%%%%%%%%%%%%%%%%%%%%%%%%%%%%%%%%%%%%%%
\subsection{The COVID-19 data}\label{sec:covid19data}

As introduced in Section \ref{sec:intro}, we are interested in exploring the relationship between the trend of epidemic situation in each country and some socio-economic factors. The data set of COVID-19 consist of the number of deaths per day from January 22, 2020, to December 15, 2020, in 190 countries and regions, as obtained from the Coronavirus Resource Center at Johns Hopkins University (https://coronavirus.jhu.edu/map.html). Due to the staggered time at which the pandemic reached individual countries, we consider a time period of 240 days, where the time 0 is the earliest day on which at least 30 deaths were reported. We take the density of the daily mortality rate, defined as the ratio of deaths per day to the total population of each country, duration a period 240 days as the response.

To present the effect of epidemic trend in each country on the overall global situation, we replace the original daily mortality rate with the relative one with respect to all countries during the period. To satisfy the requirements of the proposed method, we choose countries with the marginal distributions of covariates were compactly supported and bounded in a domain defined by the maximum and minimum of observations respectively. Then, these processes finally generate a sample of $ n=149 $ countries. The latest updated data for year 2019 of six predictors as mentioned in Section \ref{sec:intro} are collected from the World Bank (https://data.worldbank.org/indicator).

Since the raw data are relative mortality rates over daily bins, smoothing method is implemented first to construct densities. To do so, we employ the modified local linear kernel smoothers proposed by \cite{mwc97} to generate smoothing curves. The density responses $ z_{i}$, $i=1,...,149 $ are then estimated and depicted over time by Figure \ref{fg:covid}. The obvious difference in the shapes of the densities allow us to impose the latent group fixed effect in the functional additive model:
\begin{align*}
f_{i}(t)&=g_{i,0}(t)+g_{1}(t,\text{aging}_{i})+g_{2}(t,\text{beds}_{i})+g_{3}(t,\text{physicians}_{i})+g_{4}(t,\text{nurses}_{i})\\
&\qquad +g_{5}(t,GDP_{i})+g_{6}(t,\text{diabetes}_{i})+\varepsilon_{i}(t), \quad i=1,...,n,
\end{align*}
where $f_i(t)=\Psi(z_i)$ is the LQD transformation of density $z_i$.

We first use B-spline method to obtain the initial estimation $\hat{g}_{i,0}(t)$ and $\hat{g}_l(t,x_l)$, and then utilize the HAC algorithm to classify $ g_{i,0},i=1,...,n, $ with the number of clustering groups determined by the information criterion introduced in Section \ref{sec:information}.
Figure A.2 displays the values of GAIC and GBIC under the condition of different group numbers, indicating that the optimal group number is four. The members in each group are listed in Table A.2.

\begin{figure}[!ht]\centering
	\includegraphics[width=16.5cm]{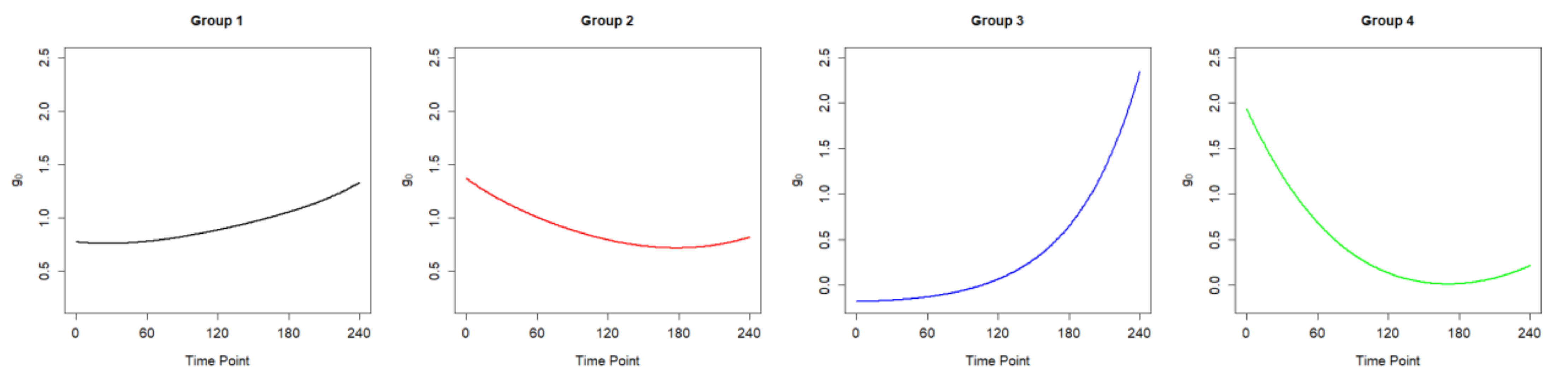}
	\caption{The estimation of latent group structures $\hat{m}_{k|4,0}(t)$.} \label{fg:covid-group}
\end{figure}

After clustering, the backfitted local linear method is employed to construct the refined estimations $\hat{m}_{k|4,0}(t)$ and $\tilde{g}_l(t,x_l)$. The estimation of latent group structures are presented by Figure \ref{fg:covid-group}, and the corresponding densities within each group are displayed by Figure \ref{fg:covid-gf}. Firstly, the trend in Group 4 is quite different with the other three, as the relative daily mortality rate increase over the time, meaning the greater proportion of mortality rate in those countries relative to the global situation during this time period. In Group 3, the daily mortality rate at the beginning is pretty high, but it dramatically declines over the time, viz., the epidemic in those countries is relatively well controlled. The trend in Group 1 also decreases over the time, but not so sharp as Group 3. Compared with the other three groups, the daily mortality rate in Group 2 fluctuates mildly.

Except for the estimation of latent groups structure, we are also interested in the influence of the selected socio-economic variables on the daily mortality rate. To quantify the contribution of each individual component function, we utilize an empirical version of the fraction of variance explained (FVE) criterion \citep{hmp19}. Specifically, the empirical FVE of the $ l$-th covariate $ x_{l} $ is defined by $ V_{l}/V_{\infty} $, where $ V_{l}=\sum_{i=1}^{n}d^{2}_{w}(z_{i},\tilde{v}_{i,0})-\sum_{i=1}^{n}d^{2}_{w}(z_{i},\tilde{v}_{i,l}) $, $ V_{\infty}=\sum_{i=1}^{n}d^{2}_{w}(z_{i},\tilde{v}_{i,0})$, with $ \tilde{v}_{i,0}=\Psi^{-1}(\tilde{g}_{i,0}(\cdot)) $, and $ \tilde{v}_{i,l}=\Psi^{-1}(\tilde{g}_{i,0}(\cdot)+\tilde{g}_{l}(\cdot,x_{i,l})) $. Then, we can find the best model by backward elimination, removing the predictor with the smallest FVE among the included predictors at each step successively. The procedure stops when the mean squared error (MSE) increases after one predictor being removed, given by $ n^{-1}\sum_{i=1}^{n}d^{2}_{w}(z_{i},\tilde{v}^{(d)}_{i}) $, where $ \tilde{v}^{(d)}_{i} $ is the fitted density of $ z_{i} $ in the $ d$-th step, and $ \tilde{v}^{(0)}_{i}=\Psi^{-1}(\tilde{g}_{i,0}(\cdot)+\sum_{l=1}^{p}\tilde{g}_{l}(\cdot,x_{i,l})) $. The MSE and FVE will be both recalculated every time a predictor is deleted. We start the backward elimination procedure at $ p=6 $, and the result shows that the variable `aging', `physicians' and `GDP' are selected in the final model, with `beds', `nurses' and `diabetes' being removed.

\begin{figure}[!ht]\centering
	\includegraphics[width=16.5cm]{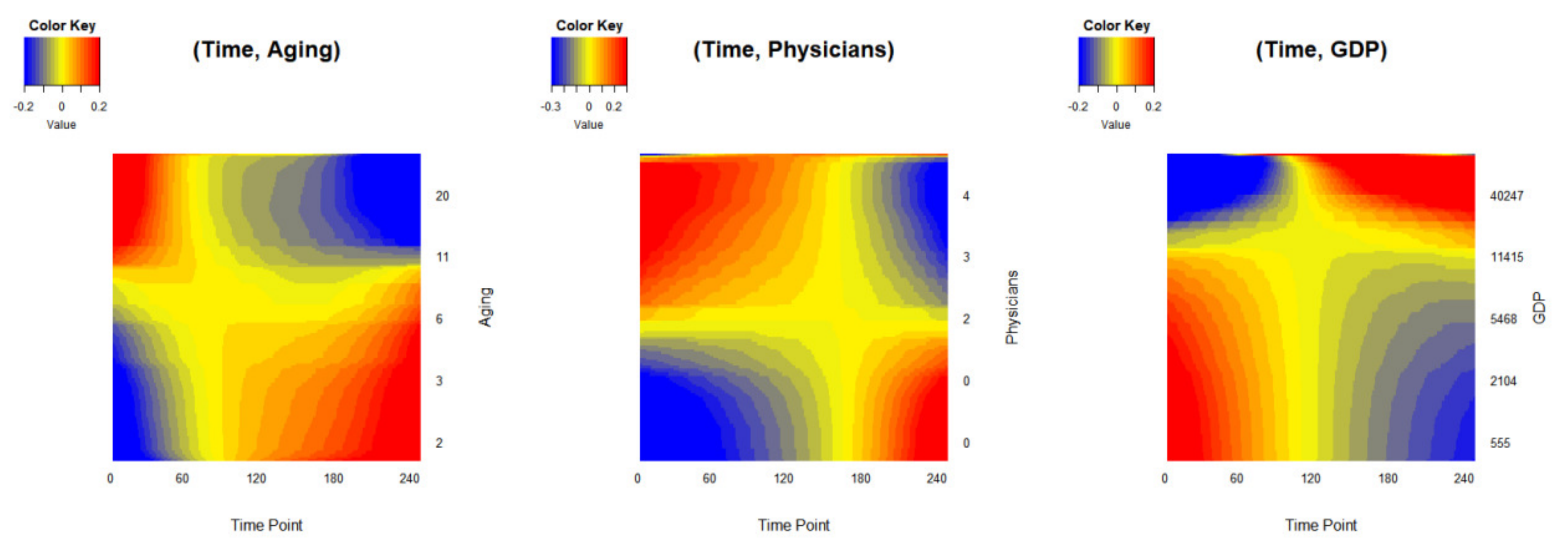}
	\caption{Heatmaps of additive component functions, where the predictors are `aging' (the percentage of population age 65 and above), `physicians' (the number of physicians per 1000 people) and GDP (the GDP per capita in United States dollar) .} \label{fg:covid-covar}
\end{figure}

Figure \ref{fg:covid-covar} demonstrates the effects of the three predictors `aging', `physicians' and `GDP' by heap maps, with FVE 45.52\%, 59.42\% and 73.98\%, respectively. The heat maps of `physicians' illustrates that the influence of `physicians' on the relative daily mortality rate changes over the time. Besides, its expressed modes are opposite for the small and large number of physicians per 1000 people. For the country with large number of physicians, the function gets to the maximum value at the early time and then minimum value at the later time, while for the country with small numbers, the pattern is opposite. Similar or opposite expressed mode can be found in the heat map of `aging' or `GDP'. 

In fact, it is not surprising to see the impacts of these predictors on the daily mortality rate. For one thing, countries with more sufficient medical supplies, more adequate medical staff, and higher domestic economic level tend to provide more effective medical treatment to reduce the incidence of death and get better control of the epidemic. For another, a handful of relevant researches reported that more various variants of the new corona virus with faster reproduction and higher infectiousness have been found in many countries and regions, which is more challenging for the immune system of the elderly. 

To evaluate the overall performance of the proposed estimations, we select three different countries from each group and draw the observed and fitted density curves in Figure A.3. On the whole, the estimated densities can well fit the observed density curves.  Meanwhile, we compare the RMSE of the pre- and the post-clustering estimators of the fitted densities, defined as $ RMSE(\tilde{v}_{i})=\frac{1}{n}\sum_{i=1}^{n}\{\frac{1}{T}\sum_{t=1}^{T}||\tilde{v}_{i}(u_{t})-z_{i}(u_{t})||^{2}_{2}\}^{1/2} $, where $ \tilde{v}_{i}=\Psi^{-1}(\tilde{g}_{i,0}(\cdot)+\sum_{l=1}^{p}\tilde{g}_{l}(\cdot,x_{i,l})) $, $ \tilde{g}_{i,0}(\cdot) $ and $ \tilde{g}_{l}(\cdot,x_{i,l}) $ are taken as pre- and post-clustering estimators respectively. The result shows that the RMSE of the pre-clustering estimations is 0.6972, while the one of the post-clustering is 0.3751, indicating that it is essential to consider the heterogeneity in the relative daily mortality rate of each country, and the identification and clustering method indeed improves the efficiency of proposed model on the analysis of COVID-19 data.

%%%%%%%%%%%%%%%%%%%%%%%%%%%%%%%%%%%%%%%%%%%%%%%%%%%%%%%%%%%%%%%%%%%%%%%%%%%%%%%%%%%%%%%%%%%%%%%%%%%%%%%%%%%%%%%%%%
\subsection{The GDP data}

GDP per capita is an effective tool to understand the macroeconomic operation of a country or region. It is one of the most important macroeconomic indicators and often used to measure the economic development. In this empirical application, we consider the model:
\begin{align*}
f_{i}(t)&=g_{i,0}(t)+g_{1}(t,\text{education}_{i})+g_{2}(t,\text{population}_{i})+g_{3}(t,\text{oriGDP}_{i})\\
&\qquad \qquad +g_{4}(t,\text{avgGDP}_{i})+\varepsilon_{i}(t), \quad i=1,...,n,
\end{align*}
where $ \text{education}_{i} $ is the the literacy rate of the $ i$-th country, namely, the percentage of educated population age 15 and above $, \text{population}_{i} $ is the total population, $ \text{oriGDP}_{i} $ is the per capital GDP of the original year, $ \text{avgGDP}_{i} $ is the average per capita GDP over 50 years.

\begin{figure}[!ht]\centering
	\includegraphics[width=8cm]{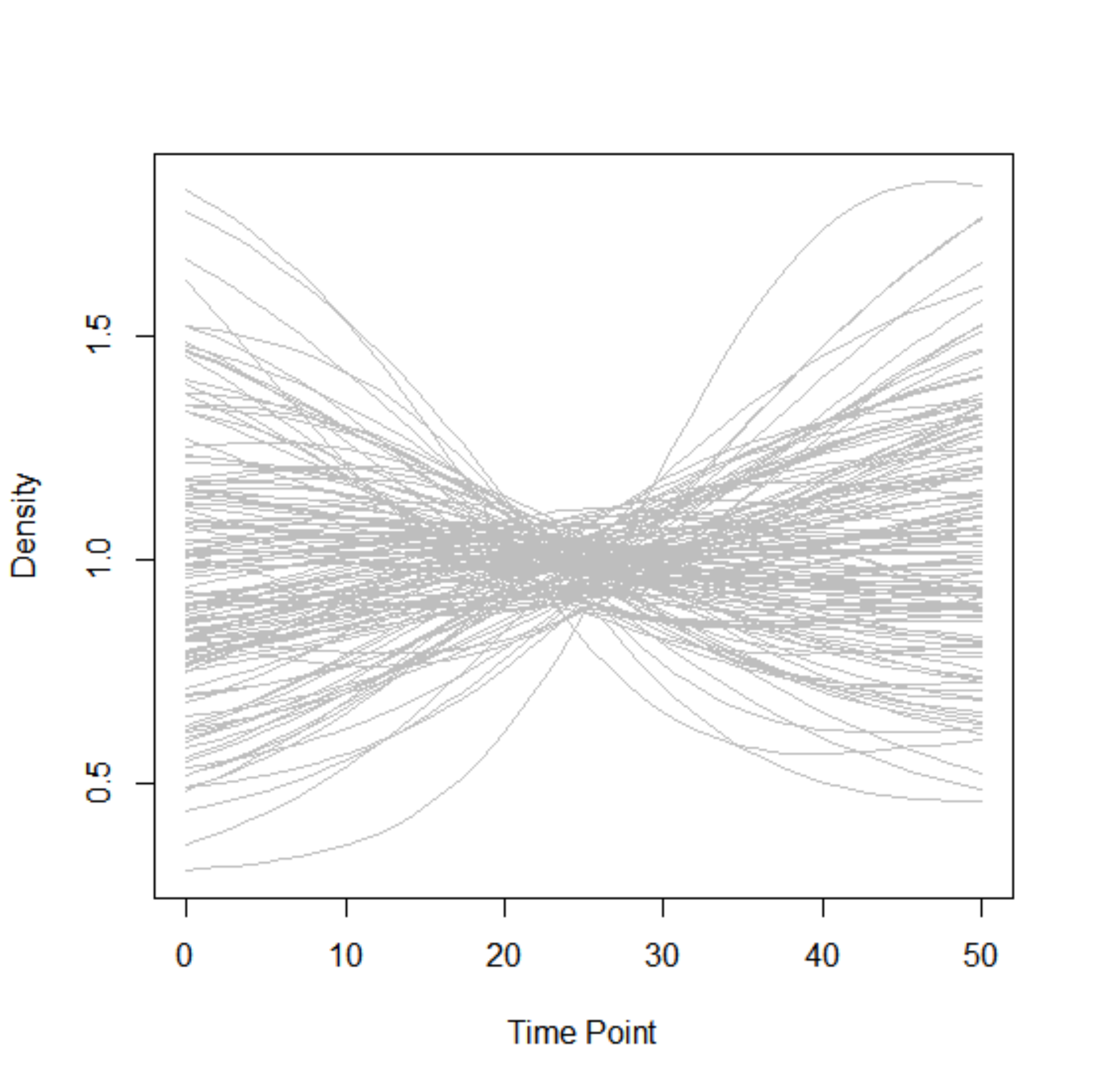}
	\caption{Densities of relative per capita GPD in 123 countries over the period of 50 years.} \label{fg:gdp}
\end{figure}

The data are obtained from the World Bank (https://data.worldbank.org) over the period 1970 to 2019. By deleting the countries with missing data, we finally have a sample of $ n=123 $ countries each with $T=50$ time points observations. With the similar procedure as Section \ref{sec:covid19data}, we obtain the estimated density of relative per capita GDP for each county to explore the economic level of each country relative to the world, which is shown in Figure \ref{fg:gdp}. It is clear that obvious difference exists in the curves among various countries.

\begin{figure}[!ht]\centering
	\includegraphics[width=13cm]{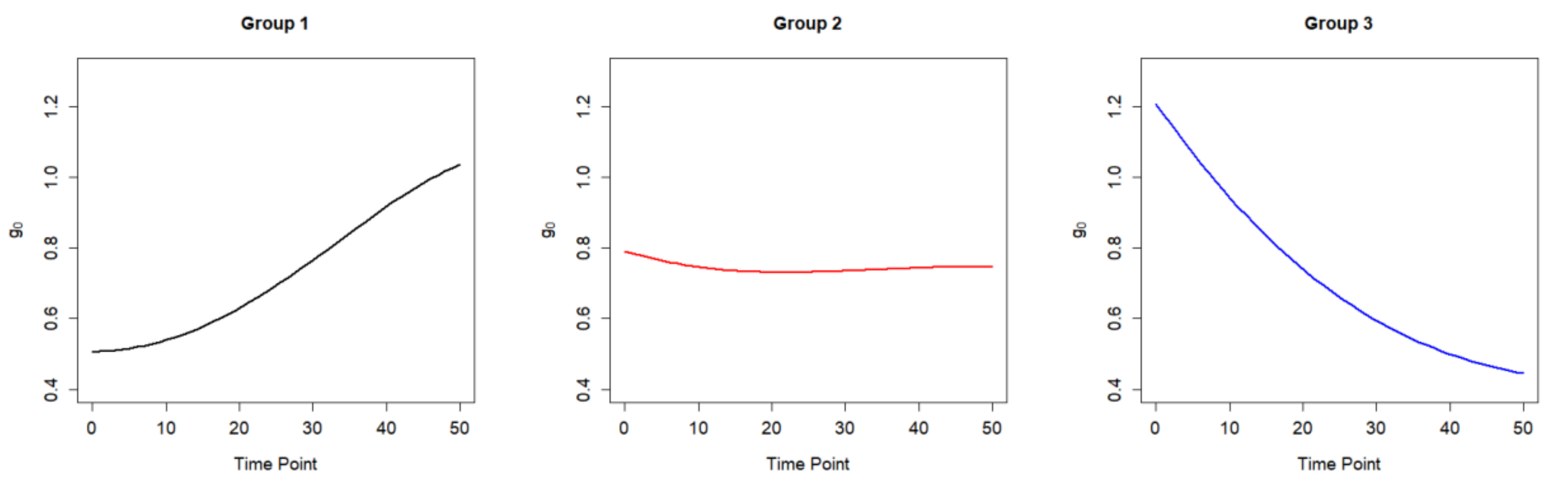}\\
	\includegraphics[width=13cm]{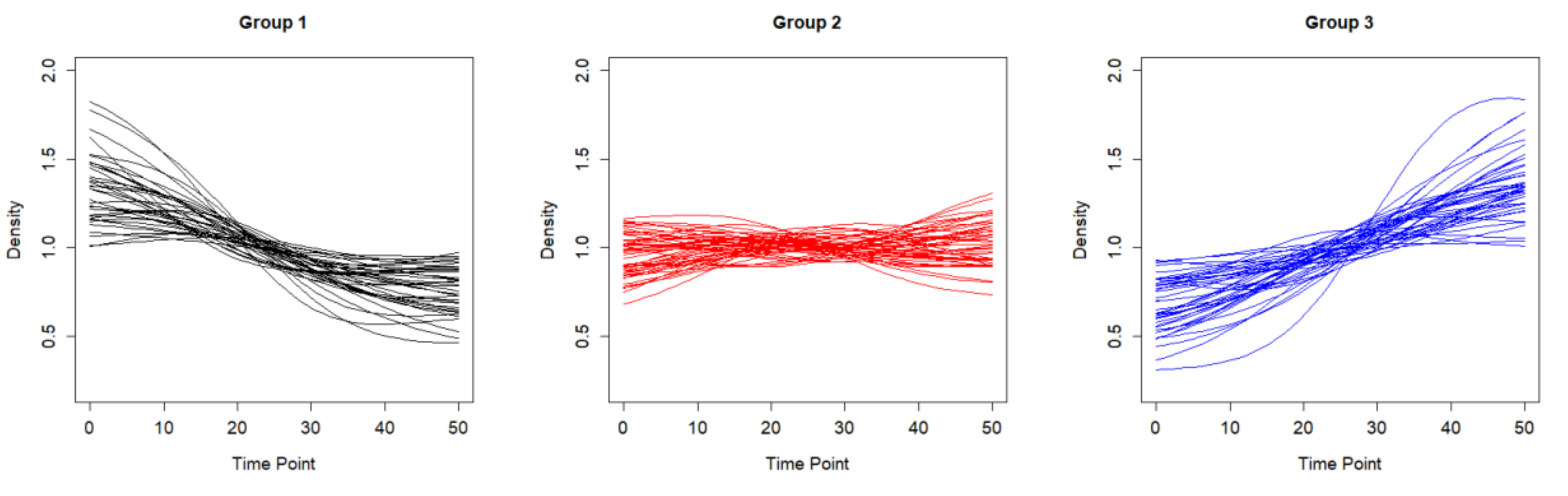}
	\caption{The estimation of latent group structures $\hat{m}_{k|3,0}(t)$ (the first row) and the densities of relative per capita GDP in each group (the second row).} \label{fg:gdp-group}
\end{figure}

Following the proposed identification and estimation procedure, based on the initial estimations $\hat{g}_{i,0}(t)$ and $\hat{g}_l(t,x_l)$, the results of HAC algorithm with GBIC and GAIC shown in Figure A.4 indicate  that $ g_{i,0}$ should be classified into three groups. The memberships in each estimated group are presented in Table A.3. Finally, post-clustering estimations $\hat{m}_{k|3,0}(t)$ and corresponding densities within each group are displayed in Figure \ref{fg:gdp-group}. 

First of all,  
relative per capita GDP in Group 3 increases over time, showing the promotion of economic influence on global development, while Group 1 has the opposite trend of decrease, indicating the decline of status in global economy. Compared with the other two groups, the effect of development in Group 2 is relatively stable.

\begin{figure}[!ht]\centering
	\includegraphics[width=16.5cm]{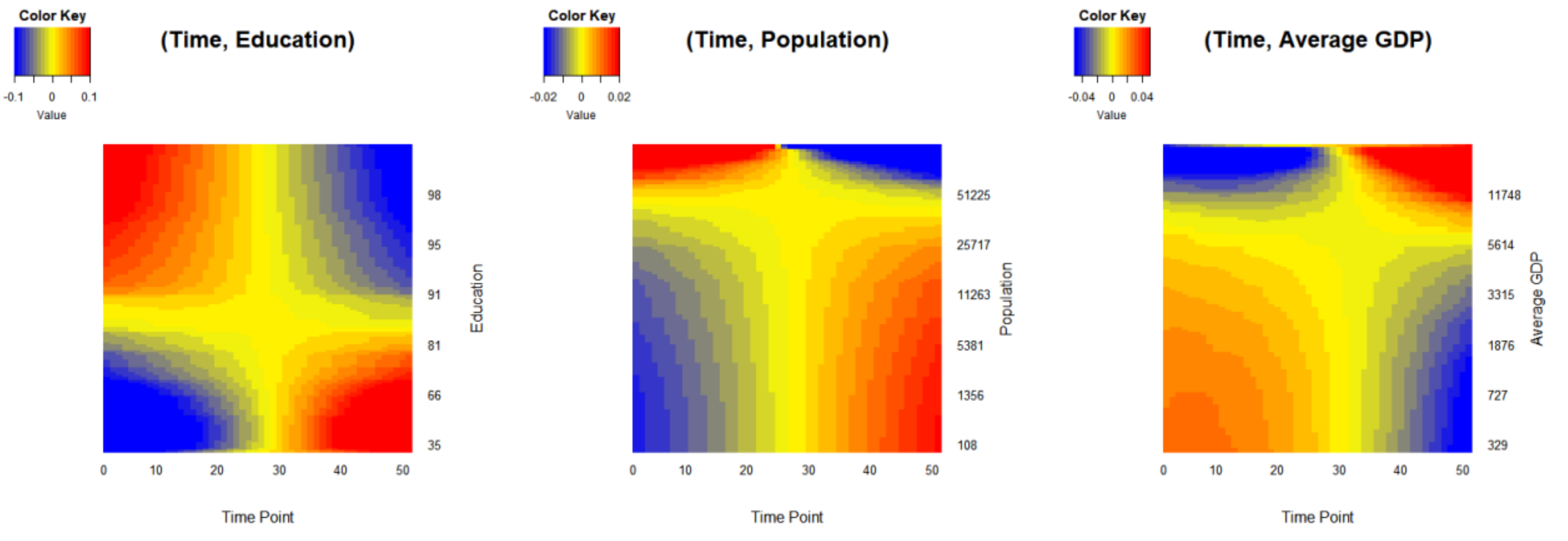}
	\caption{Heatmaps of additive component functions. The predictors are `education' (literacy rate), `population' (total population) and `average GDP' (average per capita GDP over 50 years).} \label{fg:gdp-covar}
\end{figure}

Once again backward elimination procedure with FVE criterion is implemented for model selection. The variable `oriGDP', the per capita GDP of the original year, is removed and three predictors are left for the final additive model-- `education', `population' and `average GDP'. Figure \ref{fg:gdp-covar} illustrates the impacts of these predictors via heat map, with FVE 50.75\%, 15.41\% and 30.42\%, respectively. The heat map of `education' illustrated the influence on the relative per capita GDP over the time. For the country with high literacy rate, the function gets to the maximum value at the early time and then minimum value at the later time. For the country with low rate, the function presents opposite trend. Meanwhile, the other two predictors shows similar or opposite patterns. 

To evaluate the overall performance of the proposed estimations, we select three different countries from each group and draw the observed and fitted density curves in Figure A.5. It is clear from the figure that on the whole, the estimated densities can well fit the observed density curves. Moreover, the RMSE of the pre-clustering estimations is 0.6385, while the one of the post-clustering estimations is 0.2971, indicating the necessary to consider the heterogeneity and effectiveness of the identification and clustering method of proposed model on the analysis of per capita GDP data.

%%%%%%%%%%%%%%%%%%%%%%%%%%%%%%%%%%%%%%%%%%%%%%%%%%%%%%%%%%%%%%%%%%%%%%%%%%%%%%%%%%%%%%%%%%%%%%%%%%%%%%%%%%%%%%%%%%
\section{Discussion}

With the abundance of complex data, the nature of heterogeneity and homogeneity of individual effects  existing simultaneously in the data becomes very common. On the other hand, dealing with the task of analyzing complex data that are non-Euclidean and specifically do not lie in a vector or functional space for example, the widely used density functions are increasingly popular.
To address these issues, we consider the extension of a distributional data response additive model proposed by \cite{hmp19} with heterogeneous sub populations in which the response is a distributional density function and the individual effect curves are homogeneous within a group but heterogeneous across groups, the covariates capture the common variation and share the common additive bivariate functions across subgroups. Based on the pioneer work by \cite{pm16}, a transformation approach is first applied to map density functions into a linear space. We then take the B-spline series approximating method to estimate the unknown subject-specific and additive bivariate components. The latent group structures are identified by the well known hierarchical agglomerative clustering (HAC) algorithm. We show that our method is able to identify the true latent group structures with probability approaching one. We further construct the backfitted local linear estimators for both grouped individual effect curves and the additive bivariate functions for the post-grouping model in order to improve the efficiency of the initial estimators. The asymptotic properties of the resultant estimators including the convergence rates, asymptotic distributions and the post-grouping oracle efficiency are established. The performance of the identification and clustering method is illustrated by simulation studies and two real data analysis, presenting the efficiency and validity of proposed model compared with the regular additive functional regression model without considering heterogeneity.

Many challenging problems can be addressed in future researches. In the past decades, the volume of data increases exponentially and often exceeds the available computational resources. For the massive data, functional data analysis techniques under the conditions of limited and finite samples are no longer applicable, and how to identify the latent group structures under such circumstances is a big issue for us. Meanwhile, in this paper, we consider the identification for latent group structures with only scalar covariates. In many empirical applications, functional covariates may be more preferable to present the varying correlations between predictors and responses, see \citep{chl16, lzz16, lhz17}. For this scenario, novel methods should be developed to involve both scalar and functional covariates. In addition, density functions are transformed into a linear function space via a continuous and invertible map due to the constraints for further modeling in this paper. Recently, some nice work has been done with modeling the density function response directly instead of any other transformations. See, for example, \cite{tmmhf18, pm19, clm20, jk2020a}. More relevant researches concerning this may be pursued as well in the future.

%%%%%%%%%%%%%%%%%%%%%%%%%%%%%%%%%%%%%%%%%%%%%%%%%%%%%%%%%%%%%%%%%%%%%%%%%%%%%%%%%%%%%%%%%%%%%%%%%%%%%%%%%%%%%%%%%%
%%%%%%%%%%%%%%%%%%%%%%%%%%%%%%%%%%%%%%%%%%%%%%%%%%%%%%%%%%%%%%%%%%%%%%%%%%%%%%%%%%%%%%%%%%%%%%%%%%%%%%%%%%%%%%%%%%

%%%%%%%%%%%%%%%%%%%%%%%%%%%%%%%%%%%%%%%%%%%%%%%%%%%%%%%%%%%%%%%%%%%%%%%%%%%%%%%%%%%%%%%%%%%%%%%%%%%%%%%%%%%%%%%%%%
%%%%%%%%%%%%%%%%%%%%%%%%%%%%%%%%%%%%%%%%%%%%%%%%%%%%%%%%%%%%%%%%%%%%%%%%%%%%%%%%%%%%%%%%%%%%%%%%%%%%%%%%%%%%%%%%%%
\bibliographystyle{apalike}
\bibliography{clustbib}

\begin{thebibliography}{}

\bibitem[Cardot et~al., 1999]{cfs99}
Cardot, H., Ferraty, F., and Sarda, P. (1999).
\newblock Functional linear model.
\newblock {\em Statistics \& Probability Letters}, 45(1):11--22.

\bibitem[Chen, 2019]{c19}
Chen, J. (2019).
\newblock Estimating latent group structures in time-varying coefficient panel
  data models.
\newblock {\em The Econometrics Journal}, 22(3):223--240.

\bibitem[Chen et~al., 2019]{clwz19}
Chen, J., Li, D., Wei, L., and Zhang, W. (2019).
\newblock Nonparametric homogeneity pursuit in functional-coefficient models.
\newblock {\em Working paper, University of York, New York}.

\bibitem[Chen et~al., 2020]{clm20}
Chen, Y., Lin, Z., and M\"{u}ller, H. (2020).
\newblock Wasserstein regression.
\newblock {\em arXiv:2006.09660v1}.

\bibitem[Cheng et~al., 2016]{chl16}
Cheng, M., Honda, T., and Li, J. (2016).
\newblock Efficient estimation in semivarying coefficient models for
  longitudinal/clustered data.
\newblock {\em The Annals of Statistics}, 44(5):1988--2017.

\bibitem[De~Boor, 1978]{db78}
De~Boor, C. (1978).
\newblock {\em A practical guide to splines}.
\newblock Springer-Verlag, New York.

\bibitem[Fan, 1993]{f93}
Fan, J. (1993).
\newblock Local linear regression smoothers and their minimax efficiencies.
\newblock {\em The Annals of Statistics}, 21(1):196--216.

\bibitem[Ferre and Yao, 2003]{fy03}
Ferre, L. and Yao, A. (2003).
\newblock Functional sliced inverse regression analysis.
\newblock {\em Statistics}, 37(6):475--488.

\bibitem[Ferre and Yao, 2005]{fy05}
Ferre, L. and Yao, A. (2005).
\newblock Smoothed functional inverse regression.
\newblock {\em Statistica Sinica}, 15:665--683.

\bibitem[Hall and Horowitz, 2007]{hh07}
Hall, P. and Horowitz, J. (2007).
\newblock Methodology and convergence rates for functional linear regression.
\newblock {\em The Annals of Statistics}, 35(1):70--91.

\bibitem[Han et~al., 2020]{hmp19}
Han, K., Muller, H., and Park, B. (2020).
\newblock Additive functional regression for densities as responses.
\newblock {\em Journal of the American Statistical Association},
  115(530):997--1010.

\bibitem[He et~al., 2010]{hmwy10}
He, G., Muller, H., Wang, J., and Yang, W. (2010).
\newblock Functional linear regression via canonical analysis.
\newblock {\em Bernoulli}, 16(3):705--729.

\bibitem[Hilgert et~al., 2013]{hmv13}
Hilgert, N., Mas, A., and Verzelen, N. (2013).
\newblock Minimax adaptive tests for the functional linear model.
\newblock {\em The Annals of Statistics}, 41(2):838--869.

\bibitem[Jeon and Park, 2020]{jk2020a}
Jeon, J. and Park, B. (2020).
\newblock Additive regression with hilbertian response.
\newblock {\em The Annals of Statistics}.

\bibitem[Jiang and Wang, 2011]{jw11}
Jiang, C. and Wang, J. (2011).
\newblock Functional single index models for longitudinal data.
\newblock {\em The Annals of Statistics}, 39(1):362--388.

\bibitem[Ke et~al., 2015]{kfw15}
Ke, Z., Fan, J., and Wu, Y. (2015).
\newblock Homogeneity pursuit.
\newblock {\em Journal of the American Statistical Association},
  110(509):175--194.

\bibitem[Kokoszka et~al., 2019]{kmps2019}
Kokoszka, P., Miao, H., Petersen, A., and Shang, H.~L. (2019).
\newblock Forecasting of density functions with an application to
  cross-sectional and intraday returns.
\newblock {\em International Journal of Forecasting}, 35:1304--1317.

\bibitem[Li et~al., 2017]{lhz17}
Li, J., Huang, C., and Zhu, H. (2017).
\newblock A functional varying-coefficient single-index model for functional
  response data.
\newblock {\em Journal of the American Statistical Association},
  112(519):1169--1181.

\bibitem[Luo et~al., 2016]{lzz16}
Luo, X., Zhu, L., and Zhu, H. (2016).
\newblock Single-index varying coefficient model for functional responses.
\newblock {\em Biometrics}, 72(4):1275--1284.

\bibitem[Malfait and Ramsay, 2003]{mr03}
Malfait, N. and Ramsay, J. (2003).
\newblock The historical functional linear model.
\newblock {\em Canadian Journal of Statistics}, 31(2):115--128.

\bibitem[Muller et~al., 1997]{mwc97}
Muller, H., Wang, J., and Capra, W. (1997).
\newblock From lifetables to hazard rates: the transformation approach.
\newblock {\em Biometrika}, 84(4):881--892.

\bibitem[Petersen et~al., 2019]{pcm19}
Petersen, A., Chen, C., and Muller, H. (2019).
\newblock Quantifying and visualizing intraregional connectivity in
  resting-state functional magnetic resonance imaging with correlation
  densities.
\newblock {\em Brain Connectivity}, 9(1):37--47.

\bibitem[Petersen and Muller, 2016]{pm16}
Petersen, A. and Muller, H. (2016).
\newblock Functional data analysis for density functions by transformation to a
  hilbert space.
\newblock {\em The Annals of Statisics}, 44(1):183--218.

\bibitem[Petersen and Muller, 2019]{pm19}
Petersen, A. and Muller, H. (2019).
\newblock Fr\'echet regression for random objects with euclidean predictors.
\newblock {\em The Annals of Statisics}, 47(2):691--719.

\bibitem[Ramsay, 1982]{r82}
Ramsay, J. (1982).
\newblock When the data are functions.
\newblock {\em Psychometrika}, 47(4):379--396.

\bibitem[Saha et~al., 2016]{setal16}
Saha, A., Banerjee, S., Kurtek, S., Narang, S., Lee, J., Rao, G., Martinez, J.,
  Bharath, K., Rao, A., and Baladandayuthapani, V. (2016).
\newblock Demarcate: Density-based magnetic resonance image clustering for
  assessing tumor heterogeneity in cancer.
\newblock {\em NeuroImage: Clinical}, 12:132--143.

\bibitem[Sen and Ma, 2015]{sm15}
Sen, R. and Ma, C. (2015).
\newblock Forecasting density function: Application in finance.
\newblock {\em Journal of Mathematical Finance}, 5(5):433--447.

\bibitem[Stone, 1994]{s94}
Stone, C. (1994).
\newblock The use of polynomial splines and their tensor products in
  multivariate function estimation.
\newblock {\em The Annals of Statistics}, 22(1):118--171.

\bibitem[Su et~al., 2016]{ssp16}
Su, L., Shi, Z., and Phillips, P. (2016).
\newblock Identifying latent structures in panel data.
\newblock {\em Econometrica}, 84(6):2215--2264.

\bibitem[Su et~al., 2019]{swj19}
Su, L., Wang, X., and Jin, S. (2019).
\newblock Sieve estimation of time-varying panel data models with latent
  structures.
\newblock {\em Journal of Business and Economic Statistics}, 37(2):334--349.

\bibitem[Talsk\'{a} et~al., 2018]{tmmhf18}
Talsk\'{a}, R., Menafoglio, A., Machalov\'{a}, J., Hron, K., and Fiserov\'{a},
  E. (2018).
\newblock Compositional regression with functional response.
\newblock {\em Computational Statistics \& Data Analysis}, 123(1):66--85.

\bibitem[Vogt and Linton, 2017]{vl17}
Vogt, M. and Linton, O. (2017).
\newblock Classification of nonparametric regression functions in longitudinal
  data models.
\newblock {\em Journal of the Royal Statistical Society, Series B},
  79(1):5--27.

\bibitem[Vogt and Linton, 2018]{vl18}
Vogt, M. and Linton, O. (2018).
\newblock Multiscale clustering of nonparametric regression curves.
\newblock {\em Cemmap working paper CWP08/18, University of Cambridge, UK}.

\bibitem[Yao et~al., 2005]{ymw05}
Yao, F., Muller, H., and Wang, J. (2005).
\newblock Functional linear regression analysis for longitudinal data.
\newblock {\em The Annals of Statistics}, 33(6):2873--2903.

\bibitem[Zhu et~al., 2012]{zlk12}
Zhu, H., Li, R., and Kong, L. (2012).
\newblock Multivariate varying coefficient model for functional responses.
\newblock {\em The Annals of Statistics}, 40(5):2634--2666.

\end{thebibliography}

\end{document}